\documentclass[a4paper,fleqn,usenatbib,useAMS]{mnras}

\usepackage{newtxtext,newtxmath}

\usepackage[T1]{fontenc}
\usepackage{ae,aecompl}

\usepackage{graphicx}	
\usepackage{amssymb}	
\usepackage{braket}
\usepackage{multirow}
\usepackage{threeparttable}

\usepackage[normalem]{ulem}
\usepackage[usenames]{xcolor}

\def\del#1{{}}

\def\C#1{#1}

\newcommand{\vv}{\upsilon}

\title[Cosmic ray heating in cool core clusters I]
      {Cosmic ray heating in cool core clusters -- I. Diversity of steady state solutions}

\author[S.~Jacob and C.~Pfrommer]{
Svenja Jacob$^{1, 2}$\thanks{svenja.jacob@h-its.org (SJ), christoph.pfrommer@h-its.org (CP)} and
Christoph Pfrommer$^1$\footnotemark[1]
\\
$^{1}$Heidelberg Institute for Theoretical Studies, Schloss-Wolfsbrunnenweg 35, D-69118 Heidelberg, Germany \\
$^{2}$Zentrum f\"ur Astronomie der Universit\"at Heidelberg, Astronomisches Recheninstitut, M\"onchhofsstr. 12-14, D-69120 Heidelberg, Germany\\
}

\date{Accepted XXX. Received YYY; in original form ZZZ}

\pubyear{2016}

\begin{document}
\label{firstpage}
\pagerange{\pageref{firstpage}--\pageref{lastpage}}
\maketitle

\begin{abstract}
  \noindent
  The absence of large cooling flows in cool core clusters appears to require
  self-regulated energy feedback by active galactic nuclei but the exact
  heating mechanism has not yet been identified. Here, we analyse whether a
  combination of cosmic ray (CR) heating and thermal conduction can offset
  radiative cooling. To this end, we compile a large sample of 39 cool core
  clusters and determine steady state solutions of the hydrodynamic equations
  that are coupled to the CR energy equation. We \C{find solutions} that
  match the observed density and temperature profiles for all our clusters well.
  Radiative cooling is balanced by CR heating in the cluster centres and by
  thermal conduction on larger scales, thus demonstrating the relevance of both
  heating mechanisms. Our mass deposition rates vary by three orders of
  magnitude and are linearly correlated to the observed star formation
  rates. Clusters with large mass deposition rates show larger cooling radii and
  require a larger radial extent of the CR injection function. Interestingly,
  our sample shows a continuous sequence in cooling properties: clusters hosting
  radio mini halos are characterized by the largest cooling radii, star
  formation and mass deposition rates in our sample and thus signal the presence
  of a higher cooling activity.  The steady state solutions support the
  structural differences between clusters hosting a radio mini halo and those
  that do not.
\end{abstract}




\section{Introduction}

The population of galaxy clusters can be divided into cool core (CC) and non-CC
clusters. CC clusters are characterized by low entropies and short
cooling times in the centre \citep{Cavagnolo2009, Hudson2010}. Unimpeded
radiative cooling results in cooling flows with mass deposition rates of
$1000\,\rmn{M_{\sun}\,yr^{-1}}$ \cite[see][for a review]{Peterson2006}. In
contrast, only a moderate amount of cold gas and star formation is observed,
which can be up to two orders of magnitude smaller than the predictions
\citep{Peterson2006}. In order to solve the emerging cooling flow problem an
additional heating mechanism is required that balances radiative cooling.

In the centres of CC clusters, the temperature increases with radius such
that the gas at the temperature peak functions as a heat reservoir. The
transport of heat to the centres of clusters by means of thermal conduction has
been studied in great detail \citep[e.g.][]{Bertschinger1986, Bregman1988,
  Zakamska2003, Guo2008globalstability}. Although it is possible to construct
solutions in which thermal conduction balances radiative cooling, the required
conductivity has to be fine-tuned \citep{Guo2008}. Moreover, in some clusters
such a thermal balance requires a conductivity that exceeds the theoretical
maximum, i.e.{\ }the \textit{Spitzer} value \citep{Zakamska2003}. In addition the
solutions are not locally stable on scales larger than the Field length
\citep{Kim2003, Soker2003}.  Hence, thermal conduction cannot be the sole
solution to the cooling flow problem. Nevertheless, it might still play an
important role beyond the central region at intermediate cluster radii
\citep{Voit2011}.

Another source of energy that is in principle powerful enough to balance cooling
is the feedback from the active galactic nucleus (AGN) of the brightest cluster
galaxy \citep[see e.g.][for reviews]{McNamara2007, McNamara2012}. Here, the
critical question is how to efficiently couple this energy to the intra-cluster
medium (ICM).  Various processes have been explored, including mechanical
heating by hot bubbles visible in X-ray observations
\citep[e.g.][]{Brueggen2002, Gaspari2012} as well as viscous dissipation of
sound waves \citep{Ruszkowski2004}. Additionally, the rising AGN bubbles excite
gravity modes that decay and thereby generate turbulence. Hence, dissipation of
turbulent motions is another possibility for heating the cluster gas
\citep[e.g.,][]{Zhuravleva2014}. However, recent X-ray data find a low ratio of
turbulent-to-thermal pressure in the Perseus cluster at 4 per cent, thus challenging
this scenario since low-velocity turbulence cannot spread far without being
regenerated \citep{Hitomi2016}.  This result is in line with idealized
hydrodynamical simulations, which demonstrate that the conversion of gravity
modes into turbulence is very inefficient and transfers less than 1 per cent
of the injected AGN energy to turbulence \C{\citep{Reynolds2015,
    Yang2016b}}. Moreover, all these mechanisms can only make use of one quarter
of the available enthalpy provided that the bubbles are disrupted by
Kelvin-Helmholtz instabilities within a few exponential pressure scale heights
\citep{Pfrommer2013}. The remaining enthalpy is most likely contained as
internal energy of relativistic particles and magnetic fields inside the
lobes\C{, which also modifies the interplay between jets and the cluster medium
  \citep{Sijacki2008, Guo2011}}. If the CRs are able to escape the bubbles and
fill the ICM, they can heat the cluster gas through streaming.

Streaming CRs excite Alfv\'en waves via the streaming instability
(\citeauthor{Kulsrud1969} \citeyear{Kulsrud1969}; \citeauthor{Skilling1971}
\citeyear{Skilling1971}; see also \citealt{Zweibel2013}, for a review).  The CRs
then scatter on these self-excited waves, which limits the macroscopic CR
velocity in the rest frame of the gas to approximately the Alfv{\'e}n speed
\citep[][ assuming pressure carrying CRs at GeV energies]{Wiener2013}. This
self-confinement can be very efficient since it operates on time-scales of the
order of $30\,\rmn{yr}$, which is much shorter than all other time-scales in the
cluster \citep{Wiener2013, Zweibel2013}.  The wave growth is counteracted by
damping mechanisms such as non-linear Landau (NNL) and turbulent damping
\citep{Farmer2004, Wiener2013}, which leads to an energy transfer from the CRs
to the cluster gas \citep{Wentzel1971, Guo2008}.

Importantly, CR heating allows for a self-regulated feedback loop. The CRs that
are injected by the central AGN stream outwards and heat the cluster
gas. Thereby, the CRs lose energy and become more and more dilute such that
radiative cooling eventually starts to predominate. Cooling gas can then fuel
the AGN, which launches relativistic jets that accelerate CRs. Once those escape
into the ICM, they stream again outwards and provide a source of heat.  An
important aspect are the involved time-scales: if CR heating was much slower
than the involved dynamical processes, it would not be able to efficiently heat
the gas. The free fall time-scale for a typical total density of $\rho = 9
\times 10^{-25}\rmn{~g\,cm^{-3}}$ is $\tau_\rmn{ff} = \sqrt{3 \upi/(32 G \rho)}
\approx 7 \times 10^{7}\rmn{~yr}$ \citep{KrumholtzNotes}. We compare this value
to the Alfv\'en time since CR heating is mediated by Alfv\'en waves. If we
approximate the Alfv\'en time-scale as $\tau_\rmn{A} = L/\vv_\rmn{A}$
and use a typical CR pressure scale height of $L =30\rmn{~kpc}$ and a
characteristic Alfv\'en velocity of $\vv_\rmn{A} = 200\rmn{~km~s^{-1}}$
(corresponding to a magnetic field of 10~$\umu$G and $n_{\rmn{e}} =
0.01~\rmn{cm^{-3}}$), this yields $\tau_\rmn{A} \approx1.5 \times
10^{8}\rmn{~yr}$.  Hence, the Alfv\'en time-scale is of the same order as the
free fall time-scale, which demonstrates that CR heating is sufficiently fast to
have an impact on dynamical processes. Moreover, these time-scales are in the
range of typical AGN duty cycles of a few times $10^7~\rmn{yr}$ to a few times
$10^8~\rmn{yr}$ \citep{1987MNRAS.225....1A, 2005Natur.433...45M, 2005ApJ...628..629N,
   2008MNRAS.388..625S}, which is a necessary condition for
sufficient replenishment of CRs.

For these reasons, CR heating has the potential to play a significant role in
solving the cooling flow problem \citep{Loewenstein1991, Guo2008, Ensslin2011,
  Fujita2011, Fujita2013b, Pfrommer2013}.  In particular, there exists a steady
state for spherically symmetric models, in which radiative cooling is balanced
by CR heating in the central regions and by thermal conduction further out
\citep{Guo2008}.  Unlike thermal conduction, CR heating is locally stable to
thermal fluctuations at $kT \sim 1\;\rmn{keV}$, coincident with the observed
temperature floor in some CC clusters \citep{Pfrommer2013}.  Moreover, detailed
gamma-ray and radio observations of the Virgo cluster allow for a CR
population that prevents cooling in this particular cluster
\citep{Pfrommer2013}.

Steady state solutions are a necessary condition for the viability of a
mechanism to prevent cooling flows. There are various steady state solutions for
the ICM that include different physical processes in the literature
\citep{Zakamska2003, Guo2008globalstability, Fujita2013b}.  If only the effects
of thermal conduction are considered, steady state solutions exist but the
required conductivity needs to be fine-tuned \citep{Zakamska2003}.  This
situation can be improved by including AGN feedback that is also able to reduce
the conductivity to physical values \citep{Guo2008globalstability}. However,
\citet{Guo2008globalstability} use the ``effervescent heating'' model by
\citet{Begelman2001}, which describes AGN feedback by buoyantly rising bubbles.

Motivated by the results of \citet{Guo2008} and \citet{Pfrommer2013}, we explore
steady state solutions that simultaneously take into account thermal conduction
and CR heating and discuss common characteristics of the solutions. In our
companion paper \citep{Jacob2016b}, we assess the viability of our solutions by
comparing the resulting non-thermal radio and gamma-ray emission to
observational data. This enables us to put forward an observationally supported
scenario for self-regulated feedback heating, in which an individual cluster can
either be stably heated, is predominantly cooling or is transitioning from one
state to the other.

Previous works considered at most very small cluster samples. This precludes a
sound statistical statement about the viability and applicability of the
solution to the entire CC population. Hence, we extend our analysis to a
considerably larger cluster sample. Here, we are especially interested in
clusters in which CRs have already been observed, e.g. in the form of extended
radio emission. In a subsample of CC clusters, such emission occurs as radio mini
halos (RMHs) with typical radii of a only a few hundred kpc in contrast to the
$\sim\rmn{~Mpc}$ radio halos of non-CC clusters \citep[see e.g.][for a
  review]{Feretti2012}.  Thus, we also include those clusters in the sample
selection.

This paper is structured as follows. In Section~\ref{sec:sample}, we introduce
our cluster sample and determine required properties from observations. The
governing equations of our model and our parameter choices are described in
Section~\ref{sec:model}. We discuss our results in Section~\ref{sec:discussion}
and conclude in Section~\ref{sec:conclusions}. Throughout the paper, we assume a
cosmology with $h= 0.7$, $\Omega_{\rmn{m}} = 0.3$ and $\Omega_{\Lambda} = 0.7$.

\section{Sample}
\label{sec:sample}

We analyse a total of 39 CC clusters, which are chosen from the Archive of
Chandra Cluster Entropy Profile Tables \citep[ACCEPT,][]{Cavagnolo2009}. Here, we
explain our selection criteria and perform fits to the temperature data provided
by ACCEPT\footnote{http://www.pa.msu.edu/astro/MC2/accept/}. Moreover, we correlate
the cooling time at 1~Gyr to the star formation rate (SFR) of the cD galaxy.

\subsection{Sample selection}

All clusters in our sample are CC clusters that are selected from the ACCEPT
catalogue. It provides density and temperature profiles that are obtained from
high resolution \textit{Chandra} observations that reach close to the centre of the
clusters. As in \citet{Cavagnolo2009}, we consider galaxy clusters as CC
clusters if the central value of the entropic function $K_0$ is smaller than
$30\,\rmn{keV\,cm^2}$.  For $K_0$, we use the fit values from
\citet{Cavagnolo2009}.

Ideally we would like to choose an X-ray flux limited subsample of the ACCEPT
clusters such as the extended HIghest X-ray FLUx Galaxy Cluster Sample
\citep[HIFLUGCS,][]{ReiprichHIFLUGCS}. However, this sample does not contain all clusters
with a confirmed RMH, extended diffuse radio emission in the centres of several
CC clusters with a size of up to a few hundred kpc. These sources can only be
detected if the surface brightness exceeds a limiting value that depends on the
noise properties of the observations, and effectively favours more massive
clusters at higher redshifts. Nevertheless, we include those clusters in the
sample since their non-thermal emission can be directly compared to our model.

Hence, our sample contains all 15 clusters of \citet{Giacintucci2014} that
host an RMH. Moreover, we include the CC clusters from the sample of 50 HIFLUGCS
clusters with the highest expected gamma-ray emission from pion decay that are
also in ACCEPT \citep{Pinzke2011}.  This criterion yields 23 clusters for our
sample. Moreover, we include 10 clusters with deep \textit{Chandra} data from
\citet{Vikhlinin2006}. We also add the Virgo cluster and A 2597 due to their
role in previous studies in the context of CR heating and steady state solutions
in CC clusters \citep{Zakamska2003,Guo2008globalstability,  Fujita2013b, Pfrommer2013}.  Since some of these clusters are present in more
than one of these samples, our final sample consists of 39 galaxy clusters that
are listed with some key properties in Table~\ref{tab:sample}.

\begin{figure}
  \centering
  \includegraphics{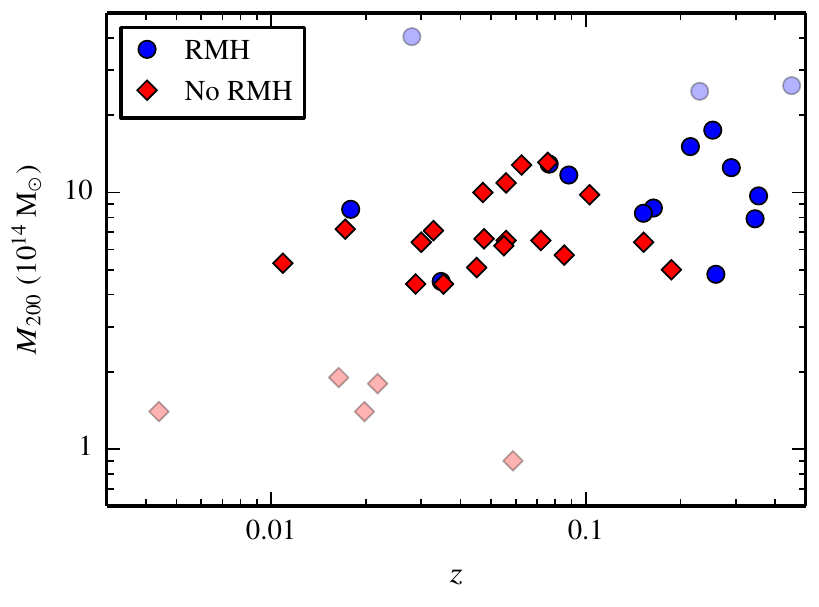}
  \caption{Cluster masses and redshifts of our sample. Clusters with an RMH
    (blue circles) have typically higher redshifts than clusters without an RMH
    (red diamonds). The majority of the clusters has comparable masses if we
    exclude individual clusters at the low and high mass end (shown with
    transparent colours).}
    \label{fig:M200Redshift}
\end{figure}

In Fig.~\ref{fig:M200Redshift}, we show cluster masses and redshifts of our
sample.\footnote{We use $M_{200}$ as an estimate for the cluster mass, which is
  the total mass contained in a sphere so that the mean density is 200 times the
  critical density $\rho_\rmn{crit} = 3 H(z)^2 / (8 \upi G)$ of the universe,
  where $H(z)$ is the Hubble function and $G$ the gravitational constant.}
Clusters that host an RMH (shown with blue circles) are the clusters with the
highest redshifts in our sample.  This is most likely due to a selection bias
associated with the limiting surface brightness effect discussed above. The
majority of our cluster sample has masses between $4\times 10^{14}$ and $2\times
10^{15}\rmn{~M_\odot}$, irrespective of whether they host an RMH or
not. However, there are five clusters without an RMH that have exceptionally low
masses (light red) and three very massive clusters with an RMH (light
blue). Where appropriate, we analyse the core sample that is (almost) unbiased
in mass and indicate the outliers only for illustrative purposes.

\begin{table*}
\caption{Cluster sample.}
\label{tab:sample}
\begin{threeparttable}
\begin{tabular}{l r r r r r r r r}
\hline
Cluster &  \multicolumn{1}{c}{$z^{\rmn{(1)}}$} &  \multicolumn{1}{c}{$K_{0}^{\rmn{(1)}}$} &  \multicolumn{1}{c}{SFR$^\rmn{(2)}$}&  \multicolumn{1}{c}{$r_{200}^{\rmn{(3)}}$}& \multicolumn{1}{c}{$r_\rmn{cool}^{\rmn{(4)}}$} & \multicolumn{1}{c}{$M_{200}^{\rmn{(5)}}$}& \multicolumn{1}{c}{$M_{200, \rmn{est}}^{\rmn{(6)}}$}& \multicolumn{1}{c}{ $\vv_\rmn{c}^{\rmn{(6)}}$}\\
&  &  \multicolumn{1}{c}{$\rmn{keV\,cm^2}$} &  \multicolumn{1}{c}{$\rmn{M_{\sun}\,yr^{-1}}$} &  \multicolumn{1}{c}{$\rmn{Mpc}$}& \multicolumn{1}{c}{$\rmn{kpc}$} &  \multicolumn{1}{c}{$10^{14}\,\rmn{M_{\sun}}$}& \multicolumn{1}{c}{$10^{14}\,\rmn{M_{\sun}}$}& \multicolumn{1}{c}{ $\rmn{km\,s^{-1}}$} \\
\hline
Centaurus	&	0.0109	&	2.25	&	0.18$^{\rmn{a}}$			&	1.67$^{\rmn{a}}$	&	10.9	&	5.3$^{\rmn{a}}$	&5.7	&	227	\\
Hydra A	&	0.0549	&	13.31	&	3.77$^{\rmn{a}}$				&	1.75$^{\rmn{a}}$	&	18.9	&	6.2$^{\rmn{a}}$	&6.5	&	326	\\
Virgo (M87)	&	0.0044	&	3.53	&	0.24$^{\rmn{a}}$			&	1.08$^{\rmn{b}}$	&	9.5		&	1.4$^{\rmn{b}}$	&3.0	&	392	\\
A 85	&	0.0558	&	12.50	&	0.61$^{\rmn{a}}$				&	2.11$^{\rmn{a}}$	&	20.0	&	10.9$^{\rmn{a}}$&12.8	&	261	\\
A 133	&	0.0558	&	17.26	&	\dots\hphantom{0}				&	1.78$^{\rmn{a}}$	&	18.7	&	6.5$^{\rmn{a}}$	&6.1	&	0	\\
A 262	&	0.0164	&	10.57	&	0.54$^{\rmn{b}}$				&	1.17$^{\rmn{a}}$	&	5.8		&	1.9$^{\rmn{a}}$	&2.4	&	273	\\
A 383	&	0.1871	&	13.02	&	5.58$^{\rmn{a}}$				&	1.54$^{\rmn{c}}$	&	32.5	&	5.0$^{\rmn{c}}$	&8.3	&	346	\\
A 496	&	0.0328	&	8.91	&	\dots\hphantom{0}				&	1.83$^{\rmn{a}}$	&	17.0	&	7.1$^{\rmn{a}}$	&5.4	&	225	\\
A 539	&	0.0288	&	22.59	&	\dots\hphantom{0}				&	1.56$^{\rmn{a}}$	&	2.5		&	4.4$^{\rmn{a}}$	&3.7	&	491	\\
A 907	&	0.1527	&	23.38	&	\dots\hphantom{0}				&	1.69$^{\rmn{c}}$	&	8.5		&	6.4$^{\rmn{c}}$	&9.3	&	250	\\
A 1644	&	0.0471	&	19.03	&	\dots\hphantom{0}				&	2.06$^{\rmn{a}}$	&	9.1		&	10.0$^{\rmn{a}}$	&6.8	&	0	\\
A 1795	&	0.0625	&	18.99	&	\dots\hphantom{0}				&	2.23$^{\rmn{a}}$	&	20.0	&	12.8$^{\rmn{a}}$&11.0	&	0	\\
A 1991	&	0.0587	&	1.53	&	\dots\hphantom{0}				&	0.89$^{\rmn{c}}$	&	17.8	&	0.9$^{\rmn{c}}$	&1.0	&	323	\\
A 2052	&	0.0353	&	9.45	&	1.4$^{\rmn{b}}$\hphantom{0}	&	1.56$^{\rmn{a}}$	&	15.0	&	4.4$^{\rmn{a}}$	&4.0	&	0	\\
A 2199	&	0.0300	&	13.27	&	0.58$^{\rmn{a}}$				&	1.77$^{\rmn{a}}$	&	13.1	&	6.4$^{\rmn{a}}$	&6.3	&	323	\\
A 2597	&	0.0854	&	10.60	&	3.23$^{\rmn{a}}$				&	1.71$^{\rmn{a}}$	&	34.1	&	5.7$^{\rmn{a}}$	&5.6	&	319	\\
A 3112	&	0.0720	&	11.40	&	4.2$^{\rmn{b}}$\hphantom{0}	&	1.78$^{\rmn{a}}$	&	19.8	&	6.5$^{\rmn{a}}$	&7.6	&	336	\\
A 3581	&	0.0218	&	9.51	&	\dots\hphantom{0}				&	1.17$^{\rmn{a}}$	&	12.9	&	1.8$^{\rmn{a}}$	&3.3	&	207	\\
A 4059	&	0.0475	&	7.06	&	0.57$^{\rmn{a}}$				&	1.79$^{\rmn{a}}$	&	7.3		&	6.6$^{\rmn{a}}$	&6.6		&	233	\\
AWM 7	&	0.0172	&	8.37	&	\dots\hphantom{0}				&	1.84$^{\rmn{a}}$	&	5.4		&	7.2$^{\rmn{a}}$	&4.6		&	424	\\
MKW3S	&	0.0450	&	23.94	&	\dots\hphantom{0}				&	1.64$^{\rmn{a}}$	&	6.6		&	5.1$^{\rmn{a}}$	&4.4		&	304	\\
MKW 4	&	0.0198	&	6.86	&	0.03$^{\rmn{a}}$				&	1.08$^{\rmn{a}}$	&	7.6		&	1.4$^{\rmn{a}}$	&2.0		&	364	\\
PKS 0745	&	0.1028	&	12.41	&	17.2$^{\rmn{b}}$\hphantom{0}	&	2.04$^{\rmn{a}}$	&	44.5	&	9.8$^{\rmn{a}}$	&12.0	&	0	\\
ZwCl 1742	&	0.0757	&	23.84	&	2.02$^{\rmn{a}}$			&	2.25$^{\rmn{a}}$	&	13.4	&	13.1$^{\rmn{a}}$	&16.0	&	0	\\
\hline
Ophiuchus$^\star$	&	0.0280 &	8.95	&	\dots\hphantom{0}			&	3.28$^{\rmn{a}}$	&	13.3		&	40.5$^{\rmn{a}}$	&16.5	&	0	\\
Perseus (A 426)	&	0.0179	&	19.38	&	34.46$^{\rmn{a}}$		&	1.95$^{\rmn{a}}$	&	34.2	&	8.6$^{\rmn{a}}$	&4.8	&	0	\\
2A 0335+096	&	0.0347	&	7.14	&	7$^{\rmn{c}}$\hphantom{.00}	&	1.58$^{\rmn{a}}$	&31.4&	4.5$^{\rmn{a}}$ &5.9	&	228	\\
A 478	&	0.0883	&	7.81	&	2.39$^{\rmn{a}}$				&	2.17$^{\rmn{a}}$	&	32.0	&	11.7$^{\rmn{a}}$	&11.0	&	358	\\
A 1835	&	0.2532	&	11.44	&	235.37$^{\rmn{a}}$				&	2.29$^{\rmn{d}}$	&	49.2	&	17.5$^{\rmn{d}}$	&25.7	&	0	\\
A 2029	&	0.0765	&	10.50	&	\dots\hphantom{0}				&	2.24$^{\rmn{a}}$	&	24.5	&	12.9$^{\rmn{a}}$	&15.7	&	531	\\
A 2204	&	0.1524	&	9.74	&	14.7$^{\rmn{b}}$\hphantom{0}	&	1.93$^{\rmn{a}}$	&	41.1	&	8.3$^{\rmn{a}}$		&6.8	&	463	\\
A 2390	&	0.2301	&	14.73	&	40.6$^{\rmn{a}}$\hphantom{0}	&	2.59$^{\rmn{d}}$	&	18.9	&	24.8$^{\rmn{d}}$	&25.9	&	0	\\
MS 1455.0+2232	&	0.2590	&	16.88	&	9.46$^{\rmn{a}}$	&	1.48$^{\rmn{c}}$	&	44.6	&	4.8$^{\rmn{c}}$		&7.1	&	569	\\
RBS 0797	&	0.3540	&	19.49	&	\dots\hphantom{0}			&	1.81$^{\rmn{c}}$	&	51.5	&	9.7$^{\rmn{c}}$		&7.8	&	250	\\
RX J1347.5-1145	&	0.4510	&	12.45	&	\dots\hphantom{0}	&	2.42$^{\rmn{c}}$	&	37.8	&	26.1$^{\rmn{c}}$	&46.3	&	0	\\
RX J1504.1-0248	&	0.2150	&	13.08	&	140$^{\rmn{d}}$\hphantom{.00}	&	2.20$^{\rmn{c}}$	&	57.0	&15.1$^{\rmn{c}}$	&19.9	&	0	\\
RX J1532.9+3021	&	0.3450	&	16.93	&	97$^{\rmn{b}}$\hphantom{.00}	&	1.70$^{\rmn{c}}$	&	51.1	&	7.9$^{\rmn{c}}$	&11.9	&	376	\\
RX J1720.1+2638	&	0.1640	&	21.03	&	\dots\hphantom{0}	&	1.87$^{\rmn{c}}$	&	32.5	&	8.7$^{\rmn{c}}$		&12.4	&	369	\\
ZwCl 3146	&	0.2900	&	11.42	&	65.51$^{\rmn{a}}$	&	2.01$^{\rmn{c}}$	&	43.8	&	12.5$^{\rmn{c}}$	&15.0	&	388	\\
\hline
\end{tabular}
\begin{tablenotes}
\item (1) Data are taken from the ACCEPT homepage \citep{Cavagnolo2009} except for Ophiuchus ($\star$), for which we use the data from \citet{Werner2016}.
\item (2) a) \citet{Hoffer2012} b) \citet{ODea2008} c) \citet{Donahue2007} d) \citet{Ogrean2010}
\item (3) a) \citet{Pinzke2011} b) \citet{Urban2011} c) $r_{\rmn{500}}$ from \citet{Lagana2013}, $r_{\rmn{200}} = r_{\rmn{500}}/0.63$ (see appendix B in \citealt{Lagana2013}), d) \citet{Ettori2010}
\item (4) We define the cooling radius $r_\rmn{cool}$ as the radius where the cooling time is $1\,\rmn{Gyr}$.
\item (5) a) \citet{Pinzke2011} b) \citet{Urban2011} c) $M_{\rmn{500}}$ from \citet{Lagana2013} d) $M_{500}$ from  \citet{Ettori2010}, for c) and d), we use $M_{200} = 200 \times 4 \pi \rho_{\rmn{crit}} r_{200}^3/3$.
\item (6) We use estimates from scaling relations, see Section~\ref{sec:modelspec}.
\end{tablenotes}
\end{threeparttable}
\end{table*}


\subsection{Temperature profiles}

For this work, it is convenient to construct continuous temperature profiles
from the ACCEPT data points to find smoother boundary conditions for the
integration of the steady state equations and to determine the maximal
temperature for each cluster (see Section~\ref{sec:model}).\footnote{Note that
  ACCEPT temperature profiles are not deprojected. While this may affect steep
  temperature profiles at small angular scales, the projection effect should not
  significantly influence our analysis and main conclusions
  \citep{Vikhlinin2006, Cavagnolo2009}. We exchange the temperature and
    density profiles for the Ophiuchus cluster and adopt a weighted average profile of the deprojected sector profiles
    from \citet{Werner2016}, which are based on significantly deeper {\em
      Chandra} data.} We also use these profiles to determine the cooling time
as a function of radius (see Section~\ref{sec:coolingtime}).

To describe the temperature profile, we use the model from \citet{Allen2001}
with the modifications introduced by \citet{Pinzke2010} in order to capture the
temperature decline at large radii,
\begin{equation}
  T = T_0 + \left(T_1 - T_0 \right)
  \left[1+ \left( \frac{r}{r_{\rmn{T}}} \right)^{-\eta} \right]^{-1}
  \left[1 + \left( \frac{r}{a r_{200}} \right)^2 \right]^{-0.32}
\end{equation}
with free parameters $T_0$, $T_1$, $r_{\rmn{T}}$ and $\eta$.  We vary the value
of $a$ for individual clusters but keep it constant during the fit. The radius
$r_{\rmn{200}}$ is defined as the radius corresponding to $M_{200}$. We use the
values for $r_{200}$ from the literature that are listed in
Table~\ref{tab:sample}. The resulting fit parameters and the maximum radius, out to
which the fit is performed, are shown in Table~\ref{tab:fitparam}.


\subsection{Cooling time profiles}
\label{sec:coolingtime}

We use the profile of the cooling time to describe the size of the central region, in which the cooling flow problem is most severe. We determine the 
cooling time as in \mbox{\citet{Donahue2005}}, such that
\begin{equation}
  \tau_\rmn{cool} (r) = 10^8\,\rmn{yr} \left[ \frac{K(r)}{10\,\rmn{keV\,cm^2}} \right]^{3/2} \left[ \frac{kT (r)}{5\,\rmn{keV}}\right]^{-1},
  \label{equ:taucool}
\end{equation}
where $k$ is the Boltzmann factor and the quantity $K(r)= kT(r)
n_\rmn{e}(r)^{-2/3}$ describes the entropic function as a function of radius
with the electron number density $n_\rmn{e}$. Here, we use the fits for $K(r)$
by \citet{Cavagnolo2009} \footnote{For Ophiuchus, we calculate $K(r)$ from our temperature and density fits (for the density see \citet{Jacob2016b}).}.

\begin{figure}
  \centering
  \includegraphics{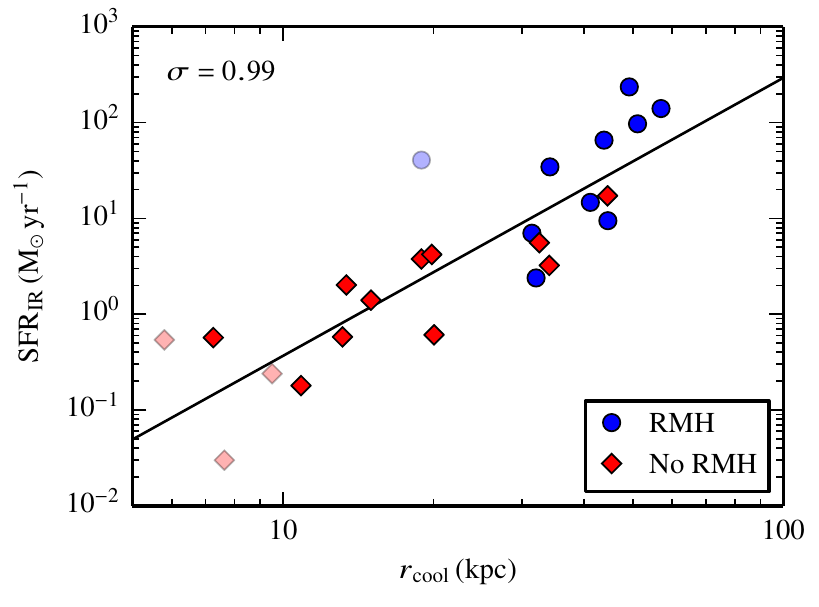}
  \caption{We compare the observed infra-red SFR to the cooling radius,
    $r_{\rmn{cool}}$, which is defined as the radius at which the gas has a
    cooling time of 1~Gyr. The larger the SFR, the larger is also the cooling
    radius. Clusters hosting an RMH (blue circles) are characterized by large
    SFRs and cooling radii. Clusters at the low- and high-mass end of our sample
    are shown with more transparent colours, indicating that the
    SFR--$r_{\rmn{cool}}$ correlation is not driven by mass.  The black line
    shows a power-law fit to the data of our core sample (shown with full
    colours) and $\sigma$ denotes the logarithmic scatter.  }
    \label{fig:rcoolvsSFR}
\end{figure}

To characterize the cooling time profile, we define the cooling radius
$r_\rmn{cool}$ as the radius where the cooling time is $1\,\rmn{Gyr}$. The
values for the cooling radius range between $2.5$ and $60\,\rmn{kpc}$ (see
Table~\ref{tab:sample} and Fig.~\ref{fig:rcoolvsSFR}), which already indicates a
substantial variance of cooling properties in our sample.  This diversity might
be connected to the differences of the inner temperature profiles
\citep{Hudson2010}.

In Fig.~\ref{fig:rcoolvsSFR}, we compare cooling radii to the observed infra-red
SFR. We take the latter from the literature as listed in
Table~\ref{tab:sample}. There is a correlation between the SFR and the cooling
radius, similar to the results from \citet{Hudson2010}. The larger the cooling
radius, the larger is also the observed SFR.  The black line shows a power-law
fit to the data with the log-normal scatter $\sigma$. We distinguish between
clusters with (blue) and without (red) RMHs. The figure demonstrates that the
clusters with the highest SFRs and largest cooling radii host RMHs and vice
versa. Moreover, this trend is not primarily driven by mass differences, which
indicates another distinction between clusters that is signalled by presence of
an RMH and may even be causally connected. We will come back to this correlation
in the analysis of the steady state solutions and in the calculation of the
non-thermal emission in our companion paper \citep{Jacob2016b}.

\section{The model}
\label{sec:model}

In the following section, we first describe the governing equations for the ICM, explain our approximations and show the resulting steady state equations. Moreover, we specify how the remaining model parameters are chosen.

\subsection{Governing equations}
\label{sec:govequ}

Adopting the simplifying assumptions that thermal and CR transport processes are
isotropic, the equations for the
conservation of mass, momentum, thermal energy and CR energy are given by
\begin{align}
\frac{\rmn{d} \rho}{\rmn{d} t} + \rho \bmath{\nabla} \bmath{\cdot } \boldsymbol{\vv}  & = 0\label{equ:cont}\\
\rho \frac{\rmn{d} \bmath{\vv}}{\rmn{d} t} &= - \bmath{\nabla} \left( P_{\rmn{th}} + P_{\rmn{cr}}\right) - \rho \bmath{\nabla} \phi\label{equ:euler}\\
\frac{\rmn{d} e_{\rmn{th}}}{\rmn{d} t} + \gamma_\rmn{th} e_{\rmn{th}} \bmath{\nabla \cdot \vv}&=  - \bmath{\nabla \cdot F}_\rmn{th}+\mathcal{H}_{\rmn{cr}} - \rho \mathcal{L} \label{equ:gasen}\\
\frac{\rmn{d} e_{\rmn{cr}}}{\rmn{d} t} + \gamma_{\rmn{cr}} e_{\rmn{cr}} \bmath{\nabla \cdot \vv} &= - \bmath{\nabla \cdot F}_{\rmn{cr}} - \mathcal{H}_{\rmn{cr}} + S_{\rmn{cr}}\label{equ:cr}
\end{align}
where $\rmn{d}/\rmn{d} t = \partial/\partial t + \bmath{\vv \cdot \nabla}$
denotes the Lagrangian derivative, $\rho$ describes the gas density,
$\bmath{\vv}$ is the mean gas velocity and $\phi$ is the gravitational
potential. The thermal pressure and energy are given by $P_{\rmn{th}}$ and
$e_{\rmn{th}}$, whereas $P_\rmn{cr}$ and $e_\rmn{cr}$ specify CR pressure and
energy. The conductive heat flux is denoted by $F_\rmn{th}$, the CR heating rate
and radiative cooling rate are denoted by $\mathcal{H}_\rmn{cr}$ and $\rho
\mathcal{L}$, respectively. $F_\rmn{cr}$ and $S_\rmn{cr}$ denote the CR
streaming flux and CR source function.\footnote{\C{In our model, we adopt the
    simplifying assumption that we can neglect turbulent heating and diffusive
    CR transport as a result of scattering off external turbulence \citep{Wiener2016}.}}

We close this set of equations with the equations of state for the thermal gas and the CRs
\begin{align}
P_{\rmn{th}} = (\gamma_\rmn{th} -1) e_{\rmn{th}}\\
P_{\rmn{cr}} = (\gamma_{\rmn{cr}} -1) e_{\rmn{cr}}
\end{align}
where $\gamma_\rmn{th} = 5/3$ is the adiabatic index for a monoatomic gas and $\gamma_{\rmn{cr}}=4/3$ is an effective adiabatic index for the CRs, for which we assume the fully relativistic value.

Thermal pressure, gas density and temperature are related by the ideal gas law
\begin{equation}
P_{\rmn{th}} = \frac{\rho k T}{\mu m_{\rmn{p}}} = \frac{\mu_{\rmn{e}}}{\mu} n_{\rmn{e}} kT
\end{equation}
with the Boltzmann factor $k$, the mean molecular weight per particle $\mu$ and per electron $\mu_{\rmn{e}}$. As in \citet{Guo2008globalstability} and \citet{Zakamska2003}, we assume a fully ionized gas with $X=0.7$ and $Y=0.28$, such that $\mu = 0.62$ and $\mu_{\rmn{e}} = 1.18$. Electron and proton number densities are related by $n_\rmn{e} = 1.19 n_\rmn{p}$.

The change in internal energy due to thermal conduction is obtained by the divergence of the conductive heat flux $\bmath{F}_\rmn{th}$, which in turn is determined by Fourier's law of conduction
\begin{equation}
\bmath{F}_\rmn{th} = - \kappa \bmath{\nabla} T. \label{equ:Fourier}
\end{equation}
The conductivity $\kappa$ is chosen as a fraction $f$ of the \textit{Spitzer} conductivity \citep{Spitzer1962}
\begin{equation}
\kappa =f \kappa_{\rmn{sp}} = 1.84 \times 10^{-5} \left(\frac{\ln \lambda}{37}\right)^{-1} f\, T^{5/2} \mathrm{erg\,s^{-1}\,K^{-7/2}\,cm^{-1}}.
\end{equation}

We describe radiative cooling in the following form:
\begin{equation}
\rho \mathcal{L} = n_{\rmn{e}}^2 \left(\Lambda_0 + \Lambda_1 T^{1/2}\right)
\end{equation}
where $\Lambda_0 = 1.2 \times 10^{-23}\,\rmn{erg\,s^{-1}\,cm^{3}}$ and
$\Lambda_1 = 1.8 \times 10^{-27}\,\rmn{erg\,s^{-1}\,cm^{3}\,K^{-1/2}}$. With
this functional form, we approximate the cooling function of
\citet{Sutherland1993} at solar metallicity, appropriate to the central regions
of CCs. In addition to the bremsstrahlung scaling with $T^{1/2}$ at high
temperatures, we include the flattening of the cooling function at temperatures
below 1 -- 2 keV as a result of cooling due to line transitions.

\C{In the self-confinement picture, the CR population propagates with a drift
  velocity relative to the rest frame of the gas. The drift velocity results
  from balancing the growth rate of the CR streaming instability and the damping
  rates due to non-linear Landau (NNL) damping \citep{Kulsrud1969} and turbulent
  damping \citep{Farmer2004}.}  NNL damping occurs, when two waves interact and
form a beat wave, which propagates with a lower phase speed than the individual
waves so that it can interact with thermal particles. Particles that move faster
than the beat wave add energy to the wave whereas particles with slower
velocities extract wave energy. Since the latter case is typical for a
Maxwellian plasma \citep{Wiener2013}, this leads to wave damping. Turbulent
damping is caused by pre-existing strong turbulence that causes Alfv{\'e}n wave
packages primarily to decay in the direction that is transverse to the magnetic
field. CRs can only resonantly interact with Alfv{\'e}n waves on their gyro
scale. If turbulence causes those waves to decay to smaller scales, the wave
growth is exponentially damped \citep{Wiener2013}.

The drift velocity for NNL damping reads for parameters relevant to the
centres of CC clusters \citep{Wiener2013}
\begin{equation}
  \label{eq:NNL}
  \vv_{\rmn{d,NNL}} = \vv_\rmn{A}
    \left( 1 + 0.002 \,
    \frac{n_{\rmn{e},-2}^{3/4}\,(kT_{2\,\rmn{keV}})^{1/4}}
         {B_{10\umu\rmn{G}}^{}\,L_{z,20\rmn{kpc}}^{1/2}\,n_{\rmn{cr,fid}}^{1/2}}
    \gamma^{(\alpha-1)/2}\right),
\end{equation}
where $n_{\rmn{e},-2}=n_{\rmn{e}}/10^{-2}\rmn{cm}^{-3}$ is the electron number
density and $kT_{2\,\rmn{keV}}=kT/2\,\rmn{keV}$ is the temperature of CCs, $\alpha$ is the spectral index of the CR proton population, $n_{\rmn{cr,fid}} = n_{\rmn{cr}}/8\times10^{-9} \rmn{cm}^{-3}$ is the
fiducial CR number density,\footnote{Adopting a power-law CR spectrum with
  spectral index $\alpha=2.4$ and low-momentum cutoff $m_\rmn{p}c/2$, the CR
  number density is $n_{\rmn{cr}}=1.3\times 10^{3} P_\rmn{cr}$.  To obtain the
  fiducial CR number density, we assume a CR-to-thermal pressure ratio of 0.1
  and values for the thermal pressure as described.}
$B_{10\umu\rmn{G}}=B/10\umu\rmn{G}$ is the magnetic field,
$L_{z,20\rmn{kpc}}=L_z/20\,\rmn{kpc}$ is the CR scale-length and $\gamma$
denotes the Lorentz factor of CRs. Similarly, if turbulent damping predominates,
the CR drift velocity is given by \citep{Wiener2013}
\begin{equation}
  \label{eq:turb}
  \vv_{\rmn{d,\,turb}} = \vv_\rmn{A}
    \left( 1 + 0.002 \,
    \frac{B_{10\umu\rmn{G}}^{1/2}\,n_{\rmn{e},-2}^{1/2}}{L_{\rmn{MHD},20\rmn{kpc}}^{1/2}\,n_{\rmn{cr,fid}}^{}}
    \gamma^{\alpha-3/2}\right),
\end{equation}
where $L_{\rmn{MHD},20\rmn{kpc}}=L_{\rmn{MHD}}/20\,\rmn{kpc}$ is the length
scale at which turbulence is excited with velocity perturbations comparable to
the Alfv\'en speed $\vv_\rmn{A}$ (i.e. with $\mathcal{M}_\rmn{A}= 1$). If
velocity perturbations are sub-Alfv\'enic then we need to extrapolate the wave
spectrum to $L_{\rmn{MHD}}$.

The drift velocity attains contributions from two modes of propagation. The
first contribution describes the advection of CRs with the frame of the
Alfv{\'e}n waves that are excited by the streaming instability and we define
this velocity as the streaming velocity.  Since CRs stream down their pressure
gradient (projected on to the local magnetic field), the streaming velocity is
given by
\begin{equation}
   \bmath{\vv}_{\rmn{st}} =
   - \rmn{sgn}(\bmath{\hat{b} \cdot \nabla} P_{\rmn{cr}}) \bmath{\vv}_{\rmn{A}}
\end{equation}
where $\bmath{\hat{b}}$ denotes the direction of the magnetic field and
$\bmath{\vv}_{\rmn{A}} = \bmath{B}/ \sqrt{4 \uppi \rho}$ is the Alfv\'en
velocity. \C{The subdominant second term in Equations~\ref{eq:NNL} and
  \ref{eq:turb} resembles the CR drift relative to the Afv\'en wave frame and
  depends on plasma conditions and the dominant damping mechanism. Neither of
  the known damping mechanisms in ionized plasma results in diffusive
  behaviour. Formally, it can be shown that \citep{Wiener2013, Wiener2016}}
\begin{equation}
  \label{eq:kappa_d}
\bmath{\nabla \cdot} \left(\kappa_\rmn{d} \bmath{\nabla} e_\rmn{cr} \right) \approx \bmath{\nabla} \bmath{\cdot} \left( e_\rmn{cr} \bmath{n} \left(\vv_\rmn{d} - \vv_\rmn{A} \right)\right)
\end{equation}
\C{with a diffusion coefficient, $\kappa_\rmn{d}$, and a normal vector pointing
  down the CR energy gradient, $\bmath{n}$. In the case of turbulent damping,
  the expression for $(\vv_{\rm D}-\vv_{\rm A})$ is independent of $\bmath{\nabla}
  P_{\rm c}$, implying that CR transport is equivalent to streaming at velocity
  $v_{\rm D}$ down the CR gradient. For NNL damping, $(\vv_{\rm
    D}-\vv_{\rm A}) \propto L_{z}^{-1/2} \propto (\bmath{\nabla} P_{\rm c})^{1/2}$. This
  is again distinct from diffusion where the flux is proportional to $\bmath{\nabla}
  P_{\rm c}$. }  The CR flux density is given by \citep[e.g.][]{Skilling1971,
  Guo2008, Pfrommer2016}
\begin{equation}
  \bmath{F}_{\rmn{cr}} = (e_{\rmn{cr}} + P_{\rmn{cr}}) \bmath{\vv}_{\rmn{st}}.
\end{equation}

Because CRs are advected with the wave frame and electric fields vanish there,
CRs cannot experience an impulsive acceleration and can only scatter in pitch
angle. Upon transforming to the rest frame of the gas, there are electric fields
associated with the propagating Alfv\'en waves representing time-varying
magnetic fields. This causes an energy transfer from the CRs to the gas, with a
volumetric heating rate \citep{Wentzel1971, Ruszkowski2016}
\begin{equation}
  \mathcal{H}_{\rmn{cr}} = - \bmath{\vv}_{\rmn{st}} \bmath{\cdot \nabla} P_{\rmn{cr}}.
\end{equation}
Since CRs stream down their gradient, this term is always positive. Therefore,
the thermal gas is invariably heated at the expense of CR energy that is used to
drive the dissipating wave field.

\subsection{Model specifications}
\label{sec:modelspec}

In order to solve the governing equations, we need to specify the gravitational
potential, the magnetic field and the CR source term.

\C{We obtain the gravitational potential by combining the results of
  \citet{Newman2013}, who find that the total mass profile in galaxy clusters is
  best described by a Navarro-Frenk-White profile (NFW, \citeyear{Navarro1997}),
  with the results by \citet{Churazov2010}. They show that the gravitational
  potential of elliptical galaxies in the cluster centres, especially Virgo, are
  well described by isothermal spheres.  Thus, we use a superposition of an NFW
  density profile $\rho_\rmn{NFW}(r) = M_{\rmn{s}}/\left[ 4 \upi r (r_{\rmn{s}}
    + r)^2\right]$ and a singular isothermal sphere. The total gravitational
  potential is then given by
\begin{equation}
  \phi(r) = \phi_{\rmn{NFW}}(r) + \phi_{\rmn{SIS}}(r) =
  - \frac{G M_{\rmn{s}}}{r} \ln\left( 1+ \frac{r}{r_{\rmn{s}}} \right)
  + \vv_{\rmn{c}}^2 \ln\left(\frac{r}{1{~\rmn{kpc}}}\right)
\end{equation}
with the scaling parameters $M_\rmn{s}$ and $r_\rmn{s}$ of the NFW profile and
the circular velocity $\vv_\rmn{c}$.  As in \citet{Zakamska2003}, we use the
peak value of the temperature profile to determine the parameters $M_{\rmn{s}}$
and $r_{\rmn{s}}$ for the NFW profile. We obtain these temperatures either from
our fits or, in rare cases, take the ACCEPT data point with the largest value of
the radial temperature profile. Then, we use equation~23 in
\citet{Afshordi2002}, which is derived from numerical studies by
\citet{Evrard1996}, to calculate $M_{200}$ and proceed as in
\citet{Zakamska2003} to obtain $M_\rmn{s}$. The estimated value for $M_{200}$ is
listed in Table~\ref{tab:sample}.  To calculate the scale radius $r_\rmn{s}$, we
use the scaling relation by \citet{Maoz1997}. }

\C{ The remaining parameter is the circular velocity, $\vv_\rmn{c}$, which
  describes the normalization of the isothermal sphere.  In the radial range
  that we consider in this work, the SIS is only dominant in the centre of the
  cluster.  Hence, we can use the normalization of the SIS to set the extent of
  this region.  To this end, we define a transition radius, $r_\rmn{t}$, at
  which the forces from the SIS and from the NFW profile are equal.  We now use
  this transition radius as a free parameter in our model, which also determines
  the normalization of the SIS (i.e. $\vv_\rmn{c}$).  }


\begin{figure}
  \centering
  \includegraphics{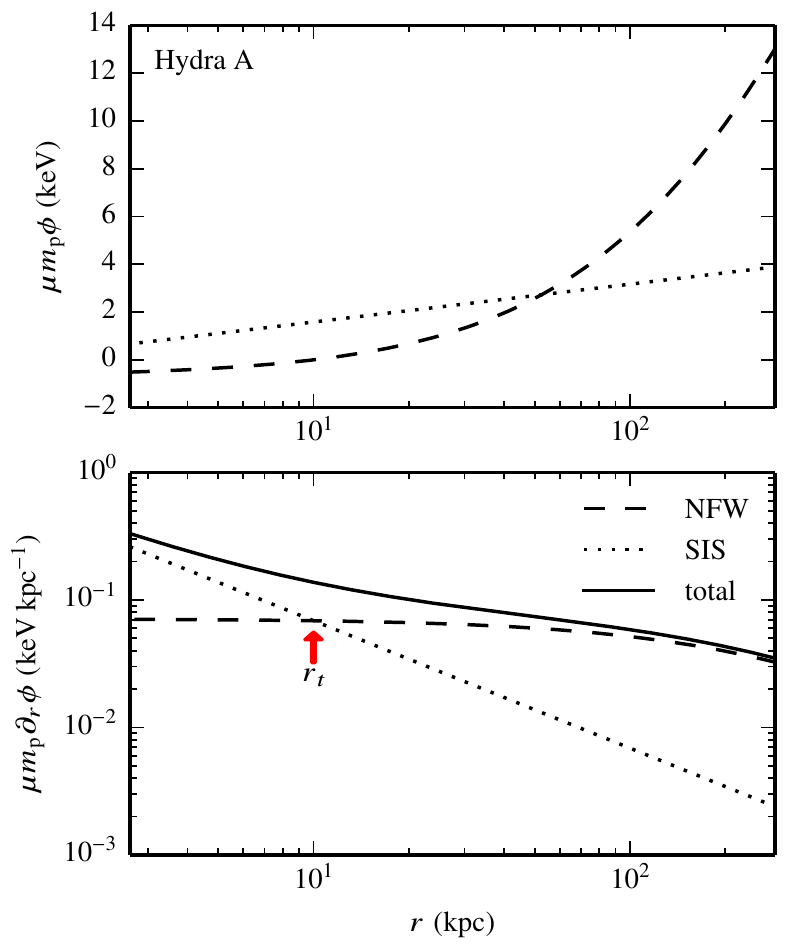}
  \caption{\C{We show our model of the gravitational potential (top) and the
    corresponding force (bottom) for the cluster Hydra~A. The potential is
    composed of a singular isothermal sphere (SIS) that dominates in the centre
    and an NFW profile at larger radii. The normalization of the SIS is chosen
    such that at the radius $r_\rmn{t}$, the forces of SIS and the NFW profile
    are equally strong. }}
    \label{fig:gravpot}
\end{figure}

\C{ Fig.~\ref{fig:gravpot} illustrates a typical example of our model of the
  gravitational potential. The top panel shows the potential and the bottom
  panel the corresponding forces in the radial range that we consider for the
  steady state solutions. The SIS dominates in the cluster centre and becomes
  subdominant to the NFW profile towards larger scales.  Even at the largest
  relevant radius for this work, the NFW profile remains the governing potential
  (while the SIS would start to predominate again at much larger radii). This
  holds for all clusters in our sample. The bottom panel highlights the
  definition of the transition radius. Here, we demand that the forces due to
  the SIS and the NFW profile are equally strong, which fixes the normalization
  of the SIS.  }

We do not model the magnetic field and its evolution explicitly but need to
parametrize it in order to calculate CR streaming. The strength and orientation
of magnetic fields in galaxy clusters are uncertain. Here, we will assume
spherical symmetry (see also Section~\ref{sec:steadystateequ}) such that it is
sufficient to model the radial magnetic field component as \citep{Vogt2005,
  Kuchar2011}
\begin{equation}
B(r) = B_0 \left( \frac{n_{\rmn{e}}}{0.01\,\rmn{cm^{-3}}} \right)^{\alpha_{B}}.
\end{equation}
We choose a magnetic field normalization $B_0=10\,\rmn{\umu G}$ and a power-law
index of $\alpha_B = 0.5$. Those values are of the same order as observed
magnetic fields in Virgo and Hydra A \citep{deGasperin2012, Kuchar2011}, but
somewhat larger than the assumption by \citet{Pfrommer2013}.  Our general
picture that CRs are injected by the central AGN implies that their number
density and pressure decrease with radius. Hence, on large scales the radial
component of the streaming velocity will be most relevant and is given by
$\vv_{\rmn{st}} = \vv_\rmn{A}$.

In the CR heating term, we also account for small-scale CR pressure fluctuations
in addition to the large-scale radial gradients. Fluctuations in the thermal
pressure are found by X-ray observations especially in the centre of the
cluster. We expect corresponding fluctuations in the CR pressure that also lead
to (non-radial) CR streaming and subsequently contribute to the CR heating
rate. Following \citet{Pfrommer2013}, we model the fluctuations as weak shocks,
such that the total CR heating rate is given by
\begin{equation}
\mathcal{H}_{\rmn{cr}} = - \bmath{\vv}_{\rmn{st}} \bmath{\cdot \nabla} P_{\rmn{cr}}
= -\vv_{\rmn{A}} \left( \frac{\rmn{d} P_\rmn{cr}}{\rmn{d} r}-\frac{5}{2}\frac{\varepsilon_{\rmn{f}} P_{\rmn{cr}}}{ r}\right)
\end{equation}
where $\varepsilon_{\rmn{f}} = 0.1$. The first term covers the large-scale
radial gradients whereas the second term describes the fluctuations.

To model CR injection, we assume that AGN feedback and thus CR injection are
triggered by accretion. Newly injected CRs are first transported inside bubbles
to a certain radius $r_{\rmn{cr}}$ and then released. The
corresponding source term is given by \citep{Guo2008}
\begin{equation}
  S_{\rmn{cr}} = - \frac{\nu \varepsilon_\rmn{cr} \dot{M} c^2}{4 \upi r_{\rmn{cr}}^3}
  \left(\frac{r}{r_{\rmn{cr}}}\right)^{-3-\nu} \left( 1- \rmn{e}^{ - (r/r_{\rmn{cr}} )^2 } \right)
\end{equation}
where $\varepsilon_\rmn{cr}$ describes how efficient the rest mass energy of the
cooling flow is converted to CR energy, $\dot{M}$ is the mass accretion rate and
\C{
$\nu$ describes the slope of the CR profile after the CRs are released into
the ambient medium at the radius $r_\rmn{cr}$. The last factor exponentially suppresses the injection of CRs at radii
that are smaller than $r_{\rmn{cr}}$.
}
We choose $\nu = 0.3$ as in the fiducial
run of \citet{Guo2008}. In the same run, \citet{Guo2008} choose $r_{\rmn{cr}} =
20\,\rmn{kpc}$ and an efficiency of $\varepsilon_\rmn{cr} =
0.003$. Nevertheless, we keep $r_{\rmn{cr}}$ and $\varepsilon_\rmn{cr}$ as free
parameters and take these values only as a first guidance.

\subsection{Steady state equations}

We obtain the steady state equations from the governing equations in
Section~\ref{sec:govequ}.  Here, we introduce the boundary conditions and the
selection criteria for the remaining parameters.

\subsubsection{Equations}
\label{sec:steadystateequ}

In order to solve Equations~\eqref{equ:cont}-\eqref{equ:cr}, we assume spherical
symmetry and only consider a steady state.  The simplified equations
are given by
\begin{align}
\dot{M} &= 4 \upi r^2 \rho \vv, \label{equ:contss}\\
\rho \vv \frac{\rmn{d} \vv}{\rmn{d} r} &= - \frac{\rmn{d}}{\rmn{d} r}\left( P_{\rmn{th}} + P_{\rmn{cr}}\right)- \rho \frac{\rmn{d} \phi}{\rmn{d} r},\label{equ:eulerss}\\
\vv \frac{\rmn{d} e_{\rmn{th}}}{\rmn{d} r} - \frac{\gamma_\rmn{th} e_{\rmn{th}} \vv}{\rho} \frac{\rmn{d} \rho}{\rmn{d} r} &=  - \frac{1}{r^2} \frac{\rmn{d}}{\rmn{d} r} \left( r^2 F_\rmn{th} \right) +  \mathcal{H}_{\rmn{cr}} - \rho \mathcal{L}\label{equ:gasenss},\\
\vv \frac{\rmn{d} e_{\rmn{cr}}}{\rmn{d} r} - \frac{\gamma_{\rmn{cr}} e_{\rmn{cr}} \vv}{\rho} \frac{\rmn{d} \rho}{\rmn{d} r} &= - \frac{1}{r^2} \frac{\rmn{d}}{\rmn{d} r} \left(r^2 F_{\rmn{cr}} \right) - \mathcal{H}_{\rmn{cr}} + S_{\rmn{cr}}, \label{equ:crss}\\
  F_\rmn{th} &= - f \kappa_{\rmn{sp}} \frac{\rmn{d} T}{\rmn{d} r}. \label{equ:fourierss}
\end{align}
Here, $\dot{M}$ denotes the mass accretion rate through each spherical shell and
$F_\rmn{cr} = \gamma_\rmn{cr} e_\rmn{cr} \vv_\rmn{A}$ the CR flux. We
numerically solve the four ordinary differential equations \eqref{equ:eulerss}
- \eqref{equ:fourierss} for the variables $\rho(r)$, $T(r)$, $F_\rmn{th}(r)$
and $P_\rmn{cr}(r)$. The fraction of the \textit{Spitzer} conductivity, $f$, is treated
as an eigenvalue of the problem, which adds a fifth differential equation
$\rmn{d}f/\rmn{d}r=0$ to the system of equations. The parameters $r_\rmn{t}$,
$\dot{M}$, $\varepsilon_\rmn{cr}$ and $r_{\rmn{cr}}$ are selected prior to the
integration according to the criteria presented in Section~\ref{sec:params}.

We choose the radius of the innermost ACCEPT data point as the inner boundary
$r_{\rmn{in}}$ of the integration. The outer radius $r_{\rmn{out}}$ is chosen
such that the temperature fits are still valid and the temperature profile is at
most at its maximum since we focus on the centre of the cluster.
To solve these five ordinary differential equations, we impose the following
five boundary conditions:
\begin{align}
\rho(r_{\rmn{in}})& = \rho_{\rmn{in}}, 	& kT(r_{\rmn{in}}) &= kT_{\rmn{in}},\\
F_\rmn{th}(r_{\rmn{in}})	 & =	F_{\rmn{th,in}},		& kT(r_{\rmn{out}}) &= kT_{\rmn{out}},\\
P_{\rmn{cr}} (r_{\rmn{in}})	&= P_{\rmn{cr,in}}.
\end{align}
Here, we use the first ACCEPT data point for $\rho_{\rmn{in}}$.\footnote{For
  two clusters (A~3581 and RX~J1504), we use the second point instead to
  avoid an increasing density profile at the centre. For the same reason, we use the maximal value in Perseus.} To determine the
temperature values $kT_{\rmn{in}}$ and $kT_{\rmn{out}}$, we generally use our
fitted temperature profiles or, if in some exceptional cases the fits are poor,
the corresponding ACCEPT data point. We prefer the smoother fits since the data
points can have substantial scatter in the outer regions of the cluster. We do
not allow any heat flux to cross the inner boundary and set $F_{\rmn{th,in}} =
0$ for all clusters. This directly implies that the temperature gradient also
vanishes there. We obtain the inner CR pressure $P_{\rmn{cr, in}}$ by solving
the steady state equations at the inner boundary. Since we want to focus on
solutions in which CR heating dominates (see also next section), we solve these
equations here without the conduction term. All boundary values are also listed
in Table~\ref{tab:params}.

\begin{table*}
\caption{Radial integration limits, boundary conditions and model parameters for our cluster sample.}
\label{tab:params}
\begin{threeparttable}
\begin{tabular}{l r r r r r r r r r r r r}
\hline
Cluster &\multicolumn{1}{c}{ $r_\rmn{in}$}&\multicolumn{1}{c}{$r_\rmn{out}$}& \multicolumn{1}{c}{$n_\rmn{e,in}^{\rmn{(1)}}$}& \multicolumn{1}{c}{$T_\rmn{in}^{\rmn{(1)}}$} &\multicolumn{1}{c}{$T_\rmn{out}^{\rmn{(1)}}$}&\multicolumn{1}{c}{$X_\rmn{cr,in}$}  & \multicolumn{1}{c}{$X_\rmn{cr, r_2}^{\rmn{(2)}}$}& \multicolumn{1}{c}{$r_\rmn{t}^{\rmn{(3)}}$ } & \multicolumn{1}{c}{$\dot{M}_{}^{\rmn{(3)}}$} & \multicolumn{1}{c}{$\varepsilon_\rmn{cr}^{\rmn{(3)}}$} &\multicolumn{1}{c}{$r_\rmn{cr}^{\rmn{(3)}}$} & \multicolumn{1}{c}{$f_{}^{\rmn{(4)}}$} \\
& \multicolumn{1}{c}{kpc}& \multicolumn{1}{c}{kpc}& \multicolumn{1}{c}{$\rmn{cm^{-3}}$}& \multicolumn{1}{c}{keV}& \multicolumn{1}{c}{keV}& && \multicolumn{1}{c}{kpc}& \multicolumn{1}{c}{$\rmn{M_{\sun}\,yr^{-1}}$}&& \multicolumn{1}{c}{kpc}\\
\hline
Centaurus	&	0.6	&	62	&	0.177	&	1.0	&	4.0	&	0.08	& 0.03\hphantom{$^{\rmn{\ast}}$}&	5	&	0.1\hphantom{0}	&	0.006	&	5	&	0.25\\
Hydra A	&	2.6	&	296		&	0.066	&	2.8	&	4.7	&	0.12	&	0.07\hphantom{$^{\rmn{\ast}}$}&	10	&	1\hphantom{.00}	&	0.006	&	10	&	0.49\\
Virgo	&	0.7	&	44		&	0.149	&	1.9	&	2.6	&	0.07	&	0.07\hphantom{$^{\rmn{\ast}}$}&	20	&	0.1\hphantom{0}	&	0.010	&	5	&	0.52\\
A 85	&	2.6	&	248		&	0.086	&	3.1	&	7.2	&	0.14	&	0.07\hphantom{$^{\rmn{\ast}}$}&	5	&	1\hphantom{.00}	&	0.010	&	10	&	0.27\\
A 133	&	2.6	&	136		&	0.041	&	2.3	&	4.5	&	0.11	&	0.05$^{\rmn{\ast}}$&	0	&	1\hphantom{.00}	&	0.003	&	10	&	0.43\\
A 262	&	0.8	&	52		&	0.037	&	1.5	&	2.4	&	0.02	&	0.02$^{\rmn{\ast}}$&	10	&	0.01				&	0.006	&	5	&	0.23\\
A 383	&	7.4	&	156	&	0.075	&	3.0	&	5.6	&	0.14	&	0.07$^{\rmn{\ast}}$&	10	&	1\hphantom{.00}	&	0.010	&	10	&	0.57\\
A 496	&	1.5	&	79		&	0.085	&	2.0	&	5.4	&	0.08	&	0.04$^{\rmn{\ast}}$&	5	&	1\hphantom{.00}	&	0.001	&	5	&	0.23\\
A 539	&	1.4	&	38		&	0.034	&	3.0	&	3.3	&	0.02	&	0.02\hphantom{$^{\rmn{\ast}}$}&	30	&	0.01	&	0.010	&	5	&	0.12\\
A 907	&	6.6	&	177		&	0.033	&	3.6	&	6.0	&	0.07	&	0.06\hphantom{$^{\rmn{\ast}}$}&	5	&	1\hphantom{.00}	&	0.003	&	10	&	0.17\\
A 1644	&	2.3	&	221		&	0.033	&	2.1	&	4.9	&	0.05	&	0.03$^{\rmn{\ast}}$&	0	&	0.1\hphantom{0}	&	0.003	&	5	&	0.24\\
A 1795	&	2.8	&	275		&	0.055	&	3.3	&	6.7	&	0.12	& 	0.07$^{\rmn{\ast}}$&	0	&	1\hphantom{.00}	&	0.006	&	10	&	0.42\\
A 1991	&	2.7	&	89		&	0.102	&	1.1	&	2.7	&	0.15	&	0.09\hphantom{$^{\rmn{\ast}}$}&	20	&	1\hphantom{.00}	&	0.003	&	10	&	0.55\\
A 2052	&	1.7	&	92		&	0.038	&	1.5	&	3.4	&	0.08	& 	0.06$^{\rmn{\ast}}$&	0	&	1\hphantom{.00}	&	0.001	&	10	&	0.32\\
A 2199	&	1.5	&	84		&	0.089	&	2.7	&	4.6	&	0.06	& 	0.06$^{\rmn{\ast}}$&	10	&	1\hphantom{.00}	&	0.003	&	10	&	0.57\\
A 2597	&	3.8	&	87		&	0.085	&	2.4	&	4.2	&	0.17	&	0.06\hphantom{$^{\rmn{\ast}}$}&	10	&	10\hphantom{.00}	&	0.001	&	10	&	0.55\\
A 3112	&	3.4	&	166		&	0.076	&	2.7	&	5.3	&	0.11	&	0.06\hphantom{$^{\rmn{\ast}}$}&	10	&	1\hphantom{.00}	&	0.006	&	10	&	0.40\\
A 3581	&	1.1	&	105		&	0.042	&	1.4	&	2.6	&	0.08	&	0.03\hphantom{$^{\rmn{\ast}}$}&	5	&	0.1\hphantom{0}	&	0.003	&	5	&	0.41\\
A 4059	&	2.3	&	140		&	0.048	&	2.1	&	4.8	&	0.07	&	0.05\hphantom{$^{\rmn{\ast}}$}&	5	&	0.1\hphantom{0}	&	0.006	&	5	&	0.19\\
AWM 7	&	0.9	&	78		&	0.086	&	2.6	&	3.8	&	0.03	&	0.03$^{\rmn{\ast}}$&	20	&	0.1\hphantom{0}	&	0.003	&	5	&	0.31\\
MKW 3S	&	2.2	&	72	&	0.036	&	3.1	&	3.6	&	0.04	&	0.03$^{\rmn{\ast}}$&	10	&	0.1\hphantom{0}	&	0.010	&	10	&	0.80\\
MKW 4	&	0.9	&	43		&	0.076	&	1.5	&	2.1	&	0.03	&	0.02$^{\rmn{\ast}}$&	20	&	0.01	&	0.010	&	5	&	0.58\\
PKS 0745	&	4.5	&	353	&	0.112	&	3.4	&	12.0	&	0.24	&	0.09$^{\rmn{\ast}}$&	0	&	10\hphantom{.00}	&	0.003	&	10	&	0.28\\
ZwCl 1742	&	3.5	&	110	&	0.045	&	3.0	&	4.6	&	0.14	&	0.09\hphantom{$^{\rmn{\ast}}$}&	0	&	1\hphantom{.00}	&	0.006	&	10	&	0.38\\
\hline
Ophiuchus$^\star$	&	1.2	&	257	&	0.234	&	1.3	&	8.8	&	0.18	& 	0.17$^{\rmn{\ast}}$&	0	&	1\hphantom{.00}	&	0.030	&	20	&	0.26\\
Perseus	&	0.9	&	114		&	0.054	&	3.2	&	6.5	&	0.05	&	0.04$^{\rmn{\ast}}$&	0	&	1\hphantom{.00}	&	0.006	&	20	&	0.37\\
2A 0335	&	1.6	&	148	&	0.120	&	1.6	&	4.4	&	0.22	&	0.07\hphantom{$^{\rmn{\ast}}$}&	5	&	1\hphantom{.00}	&	0.010	&	10	&	0.67\\
A 478	&	4.1	&	232		&	0.108	&	3.1	&	6.7	&	0.17	&	0.07$^{\rmn{\ast}}$&	10	&	1\hphantom{.00}	&	0.006	&	5	&	0.58\\
A 1835	&	9.4	&	590		&	0.117	&	4.2	&	11.8	&	0.21	&	0.16$^{\rmn{\ast}}$&	0	&	10\hphantom{.00}	&	0.010	&	20	&	0.32\\
A 2029	&	3.4	&	264		&	0.128	&	4.2	&	8.5	&	0.08	&	0.07$^{\rmn{\ast}}$&	20	&	10\hphantom{.00}	&	0.001	&	10	&	0.19\\
A 2204	&	6.5	&	124		&	0.133	&	3.3  &	9.8	&	0.14	&	0.09$^{\rmn{\ast}}$&	20	&	1\hphantom{.00}	&	0.030	&	20	&	0.15\\
A 2390	&	8.7	&	305		&	0.065	&	4.4	&	11.9	&	0.11	&	0.08$^{\rmn{\ast}}$&	0	&	10\hphantom{.00}	&	0.003	&	20	&	0.07\\
MS 1455	&	9.5	&	162	&	0.082	&	3.6	&	5.0	&	0.13	&	0.08\hphantom{$^{\rmn{\ast}}$}&	30	&	10\hphantom{.00}	&	0.003	&	20	&	0.69\\
RBS 797	&	6.2	&	241	&	0.096	&	4.4	&	9.9	&	0.22	&	0.11$^{\rmn{\ast}}$&	5	&	10\hphantom{.00}	&	0.010	&	20	&	0.43\\
RX J1347	&	14.3	&	186	&	0.128	&	7.7	&	17.5	&	0.24	&	0.15$^{\rmn{\ast}}$&	0	&	10\hphantom{.00}	&	0.030	&	20	&	0.14\\
RX J1504	&	8.3	&	537	&	0.105	&	4.6	&	10.0	&	0.20	&	0.16$^{\rmn{\ast}}$&	0	&	10\hphantom{.00}	&	0.010	&	20	&	0.72\\
RX J1532	&	11.6	&	361	&	0.093	&	4.1	&	7.1	&	0.23	&	0.10\hphantom{$^{\rmn{\ast}}$}&	10	&	10\hphantom{.00}	&	0.010	&	20	&	0.94\\
RX J1720	&	6.7	&	287	&	0.076	&	4.4	&	7.3	&	0.25	&  0.17\hphantom{$^{\rmn{\ast}}$}&	10	&	10\hphantom{.00}	&	0.010	&	20	&	0.23\\
ZwCl 3146	&	10.3	&	238	&	0.104	&	3.8	&	8.3	&	0.24	& 	0.11\hphantom{$^{\rmn{\ast}}$}&	10	&	10\hphantom{.00}	&	0.010	&	20	&	0.29\\
\hline
\end{tabular}
\begin{tablenotes}
\item (1) Data are taken from the ACCEPT homepage \citep{Cavagnolo2009} except for Ophiuchus ($\star$), for which we use the data from \citet{Werner2016}.
\item (2) Values represent our steady state solutions that are evaluated at
  $r_\rmn{cool}$ (no asterisk) or at the radius where CR and conductive heating
  are equal (denoted by an asterisk, $\ast$); see
  Section~\ref{sec:steadystatesolutions}.
\item (3) These parameters are chosen prior to the integration and kept fixed.
\item (4) The \textit{Spitzer} fraction for thermal conductivity ($f$) is treated as an eigenvalue.
\end{tablenotes}
\end{threeparttable}
\end{table*}

\subsubsection{Parameters}
\label{sec:params}

The steady state equations still contain four free parameters: $r_\rmn{t}$,
$\dot{M}$, $\varepsilon_\rmn{cr}$ and $r_{\rmn{cr}}$. In order to simplify the
integration, we specify these parameters before solving the equations. The
values of these parameters can have a significant impact on the solutions.  We
use this freedom to obtain physical solutions and to focus on CR
heating. Therefore, we scan a grid in the parameter space, allowing for all
parameter combinations.

The transition radius $r_\rmn{t}$ that links the NFW profile at large scales to
that of an isothermal sphere in the centre, assumes values of $0$, $5$, $10$,
$20$ and $30\,\rmn{kpc}$. These values allow for a pure NFW profile and reach
the maximum size of the central galaxy whose potential might well be described
by an isothermal sphere \citep{Churazov2010}. We adopt a maximum value for the
accretion rate of $10\,\rmn{M_{\sun}\,yr^{-1}}$ and decrease its value by
factors of ten because we aim at solutions without large cooling flows. 
The efficiency of transforming accreted mass into CR energy,
$\varepsilon_\rmn{cr}$, varies between $0.001$, $0.003$, $0.006$, $0.01$ and
$0.03$, with a fiducial value of $0.003$ \citep{Guo2008}. The radius
$r_{\rmn{cr}}$, which describes how far CRs are transported into the ICM by
bubbles, varies between $5$, $10$ and $20\,\rmn{kpc}$.

From this set of parameters, \C{we choose solutions} that fulfil the following
criteria. We only accept physical solutions, for which the required fraction of
the \textit{Spitzer} conductivity is smaller than unity. The theoretically favoured value
is $f \sim 0.3$ or even lower \citep[e.g.,][]{Narayan2001,
  Komarov2016}. Moreover, we only accept solutions, whose density and
temperature profiles agree well with observations. From the resulting set of
solutions, we select those which maximize CR heating. In order to meet
constraints from the literature, we require that the central CR-to-thermal
pressure value is smaller than $0.3$ \citep{Churazov2010}. Note however, that
the required CR pressure also depends on the magnetic field: the larger the
magnetic field, the less CRs are necessary to achieve the same amount of
heating. As a last criterion, we favour solutions with decreasing CR-to-thermal
pressure profiles towards larger radii.

In conclusion, we select parameters that reproduce X-ray observations and make
CRs the dominant heat source for a large radial range. The chosen parameters are
listed in Table~\ref{tab:params}. 

\begin{figure*}
  \centering
  \includegraphics{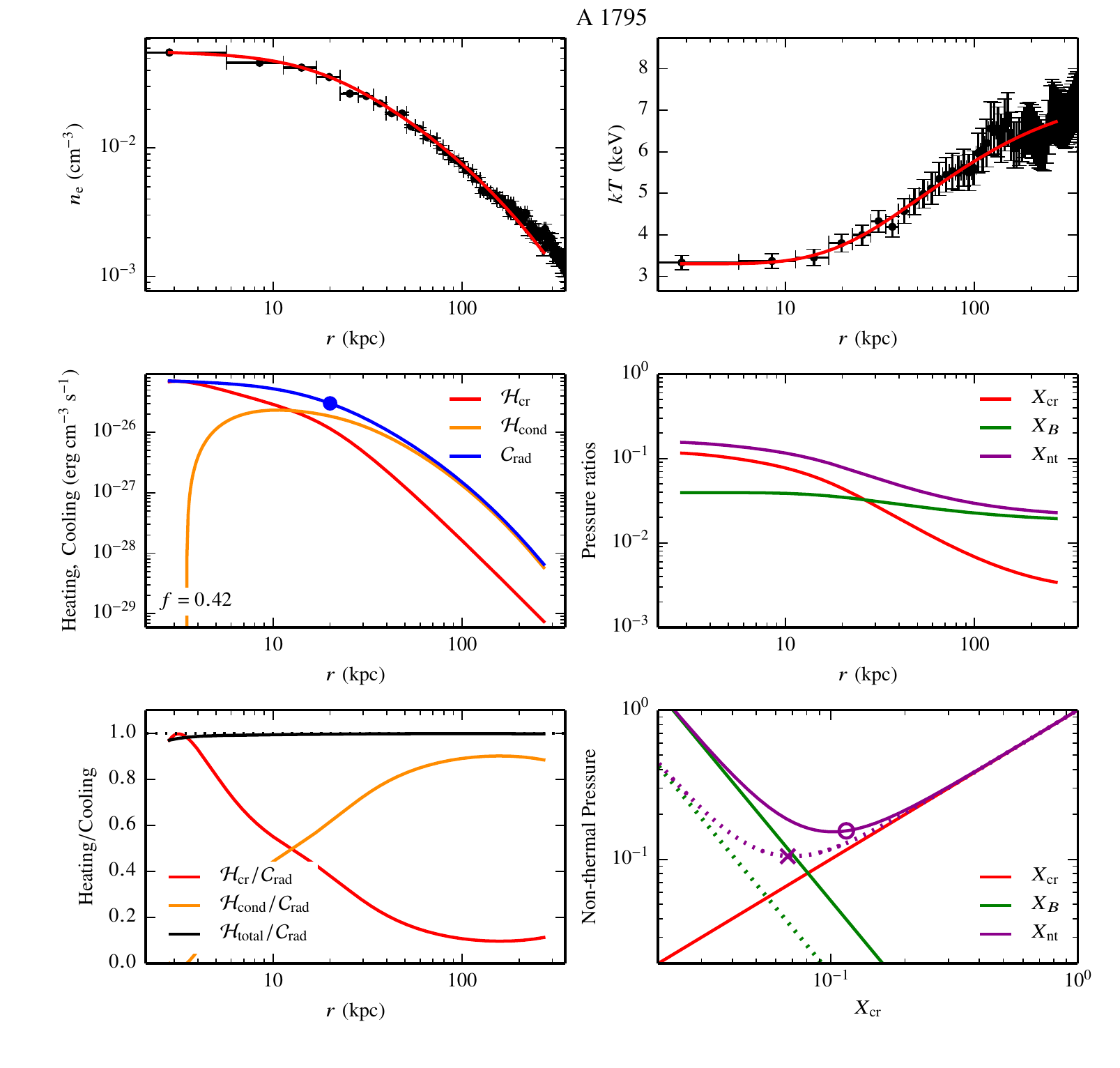}
  \caption{Dissecting the steady state solution for the cluster A 1795.
    \textit{Top.} Electron number density and temperature profiles. The data
    points are taken from ACCEPT.  \textit{Middle left.}  Cooling and heating
    rates as a function of radius; the filled circle corresponds to the location
    of the cooling radius. CR heating dominates in the centre of the cluster and
    thermal conduction becomes more important at larger radii.  \textit{Bottom
      left.}  Ratio between the different heating rates to the cooling
    rate. Note that the total heating-to-cooling ratio (black) is less than
    unity, indicating a small net cooling that causes mass accretion towards the
    centre.  \textit{Middle right.} Ratios of CR-to-thermal pressure
    $X_\rmn{cr}$, magnetic-to-thermal pressure $X_B$, as well as total
    non-thermal-to-thermal pressure $X_\rmn{nt}$ as a function of radius.
    \textit{Bottom right.} We show the relation between CR pressure and magnetic
    fields if CR heating balances radiative cooling. The smaller the magnetic
    field, the more CRs are required and vice versa. The solid lines show the
    relation at the inner boundary of our solution, the dotted lines correspond
    to the radius at which CR and conductive heating are equal, as indicated by
    the cross. The symbols represent the values of the steady state solutions.
  }
    \label{fig:steadystatedemo1}
\end{figure*}

\begin{figure*}
  \centering
  \includegraphics{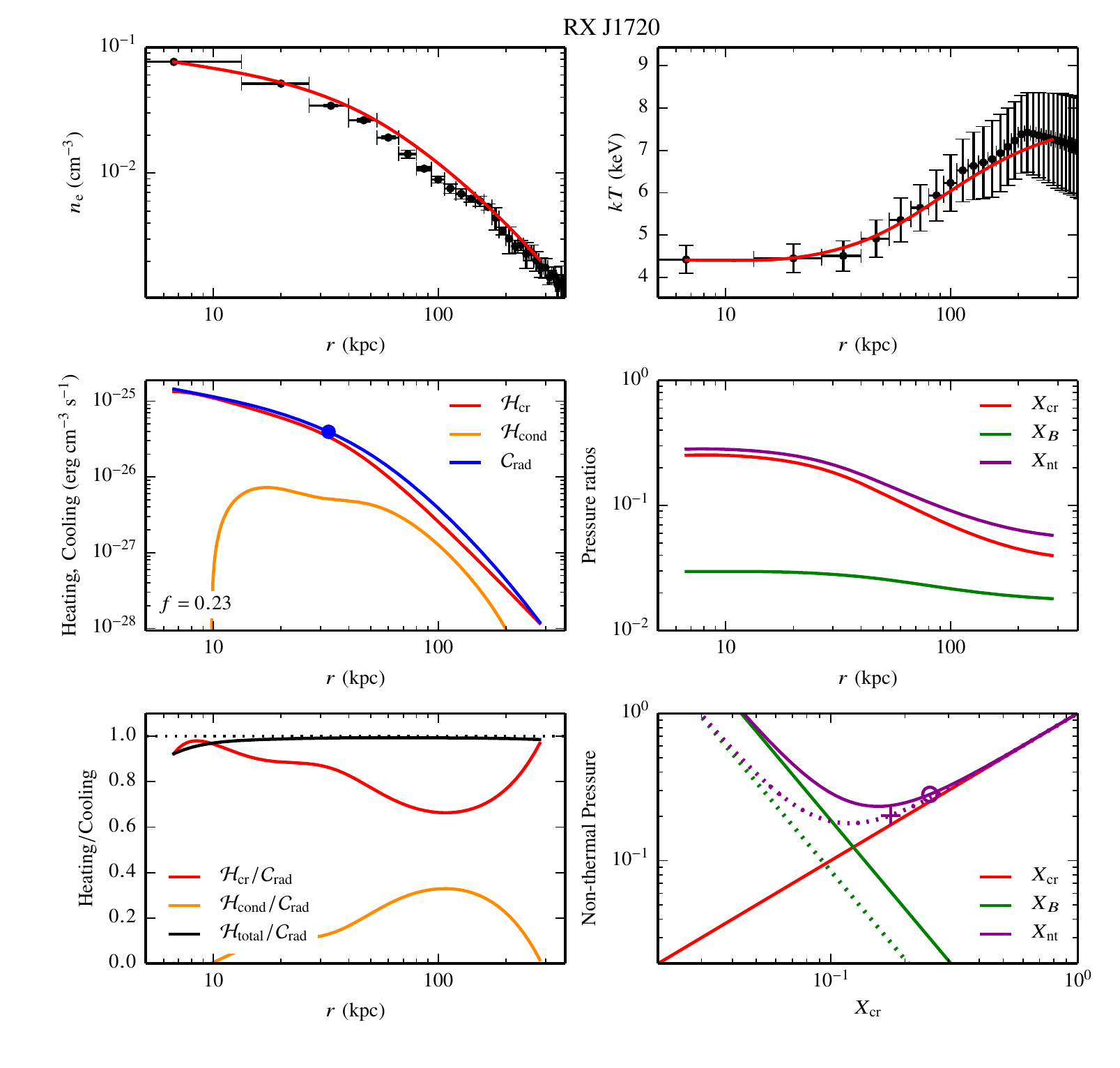}
  \caption{Same as in Fig.~\ref{fig:steadystatedemo1} but for RX J1720 which
    hosts an RMH. The plus sign in the bottom-right plot indicates that the
    dotted lines correspond to the cooling radius.}
    \label{fig:steadystatedemo2}
\end{figure*}

\section{Results}
\label{sec:discussion}

We present and discuss the steady state solutions of our fiducial model with CR
and conductive heating for two representative clusters and refer the reader to
Appendix~\ref{sec:app} for the remaining part of the sample.  In order to
understand the impact of CR heating, we additionally explore a straw man's model
with conductive heating only. We close this section by analysing the parameter
values for our fiducial model that give the best-fitting steady state solution with
CRs as the dominant heat source.

\subsection{Steady state solutions}
\label{sec:steadystatesolutions}

In Figs.~\ref{fig:steadystatedemo1} and \ref{fig:steadystatedemo2}, we show
different aspects of the steady state solutions for two example clusters, A 1795
and RX J1720. RX J1720 hosts a confirmed RMH while A 1795 does not.

The top panels of Figs.~\ref{fig:steadystatedemo1} and
\ref{fig:steadystatedemo2} show the steady state solutions for the electron
number density and temperature. The data points are taken from the ACCEPT
sample. Data and steady state solutions agree well for both clusters. Part of
the remaining discrepancies could be due to deviations from our assumptions of a
steady state or spherical symmetry. This assumption does not account for
features such as the observed bubbles in the central regions of many CC clusters
\citep[e.g.][]{Birzan2004}. In addition to that, inaccuracies especially in the
description of the gravitational potential can have large effects on the
resulting density and temperature profiles. Still, we see that the considered
physical processes \C{admit steady state solutions} that agree well with the
observed (azimuthally averaged) thermodynamic profiles. 

While this is a necessary requirement for a plausible heating mechanism, it is
not sufficient due to \C{potential local and global instabilities. We do not
  carry out stability analyses for our particular set up. However,
  \citet{Pfrommer2013} demonstrates local stability of CR heating at
  temperatures $kT\gtrsim3$~keV, around 1~keV and finds further islands of
  stability at locations of cooling line complexes in the cooling function. A
  global stability analysis for CR heating has been carried out by
  \citet{Fujita2013b}, who could not find any unstable
  modes. \citet{Guo2008globalstability} show that a combination of thermal
  conduction and AGN heating can also be globally stable if the AGN feedback is
  strong enough, thus providing circumstantial evidence that our solutions are
  likely sufficiently stable on time-scales relevant for reaching self-regulated
  heating.}

In order to scrutinize the steady state solutions further, we show the relative
merits of CR (red) and conductive heating (orange) in comparison to radiative
cooling (blue) in the middle-left panels of Figs.~\ref{fig:steadystatedemo1} and
\ref{fig:steadystatedemo2} and present both heating rates in units of the
cooling rate in the bottom-left panels of these figures. For the chosen set of
parameters, CR heating dominates in the centres of the clusters. In A~1795
thermal conduction takes over at $\sim 10\,\rmn{kpc}$, whereas in RX J1720 CR
heating stays dominant over the entire radial range that we consider here. The
latter is less typical for the complete sample since thermal conduction usually
starts to dominate in the intermediate parts of the cluster, which demonstrates
its importance at those radii (see Appendix~\ref{sec:app}). In the middle-left
panels of Figs.~\ref{fig:steadystatedemo1} and \ref{fig:steadystatedemo2}, we
also indicate the required fraction of the \textit{Spitzer} conductivity ($f=0.42$ and
$0.23$ for A 1795 and RX J1720, respectively) that will be discussed further in
the next section. The solid black line in the bottom-left panels of these
figures shows the total heating rate in units of the cooling rate. CR and
conductive heating do not exactly add up to the cooling rate because the mass
flux (which is by construction constant in each radial shell) and hence the
central mass deposition rate are non-zero. As a result, the energy equation contains advection
and adiabatic terms that do not vanish (see
Equation~\ref{equ:gasenss}). Moreover, these terms lead to radial variations of
the heating-to-cooling rate ratio.

In the middle-right panel of Figs.~\ref{fig:steadystatedemo1} and
\ref{fig:steadystatedemo2}, we show radial profiles of the ratio of
CR-to-thermal pressure, $X_\rmn{cr} = P_\rmn{cr}/P_\rmn{th}$, and the
magnetic-to-thermal pressure ratio, $X_B = B^2/\left(8 \upi
P_\rmn{th}\right)$. In both clusters, the CR-to-thermal pressure ratio peaks in
the centre and falls off to larger radii as expected for a CR population that is
injected by a central AGN. The maximal CR pressure ratio in A~1795 is
$X_\rmn{cr, max} \approx 0.10$ and $X_\rmn{cr, max} \approx 0.25$ in RX J1720.
Note that the CR-to-thermal pressure ratio is almost constant in the central
regions of the clusters and only starts to fall off rapidly beyond the cooling
radius. Such a constant pressure ratio is theoretically expected from a steady
state where CR heating balances cooling \citep{Pfrommer2013}. This can be seen
by estimating the energy per unit volume that is transferred from the CRs to the
thermal gas in steady state,
\begin{equation}
  \label{eq:constXcr}
  \Delta \varepsilon_\rmn{th} =
  -\tau_\rmn{A} \bmath{\vv}_\rmn{st} \bmath{\cdot \nabla} P_\rmn{cr}
  \approx P_\rmn{cr} = X_\rmn{cr} P_\rmn{th},
\end{equation}
where $\tau_\rmn{A} = \delta l / \vv_\rmn{A}$ denotes the Alfv\'en crossing time
over a CR pressure gradient length $\delta l$. 

Fig.~\ref{fig:steadystatedemo1} also shows that the CR pressure is larger than
the magnetic pressure in the centre of A~1795.  \C{ At larger radii, the CR
  pressure decreases faster than the magnetic pressure such that the later
  starts to dominate at a radius of $r\gtrsim 30\rmn{~kpc}$.} We see the same
trends in RX J1720 (Fig.~\ref{fig:steadystatedemo2}) but there the CR pressure
is generally larger and thus stays dominant at all radii.

In this section, we have restricted the discussion to two example clusters but
the applicability of our model to the whole sample is a key result of this
paper. We present plots similar to Figs.~\ref{fig:steadystatedemo1} and
\ref{fig:steadystatedemo2} for the other clusters of our sample in
Appendix~\ref{sec:app}. The density and temperature profiles of the steady state
solutions in our sample agree well with observations. Radiative cooling is
typically balanced in the centre by CR heating and in the intermediate parts of
the cluster, closer to the temperature peak, by conductive heating. While the
agreement of model and observed thermodynamic variables (such as density and
temperature) is a necessary requirement for a viable model, the predicted CR and
magnetic pressure values must not conflict with any other observational
data. This mainly concerns dynamical potential estimates and non-thermal
radio and gamma-ray observations of these clusters. We will return to this point
in our companion paper \citep{Jacob2016b}.

\subsection{Non-thermal pressure constraints}
\label{sec:NTpressure}

\begin{figure*}
  \centering
  \includegraphics{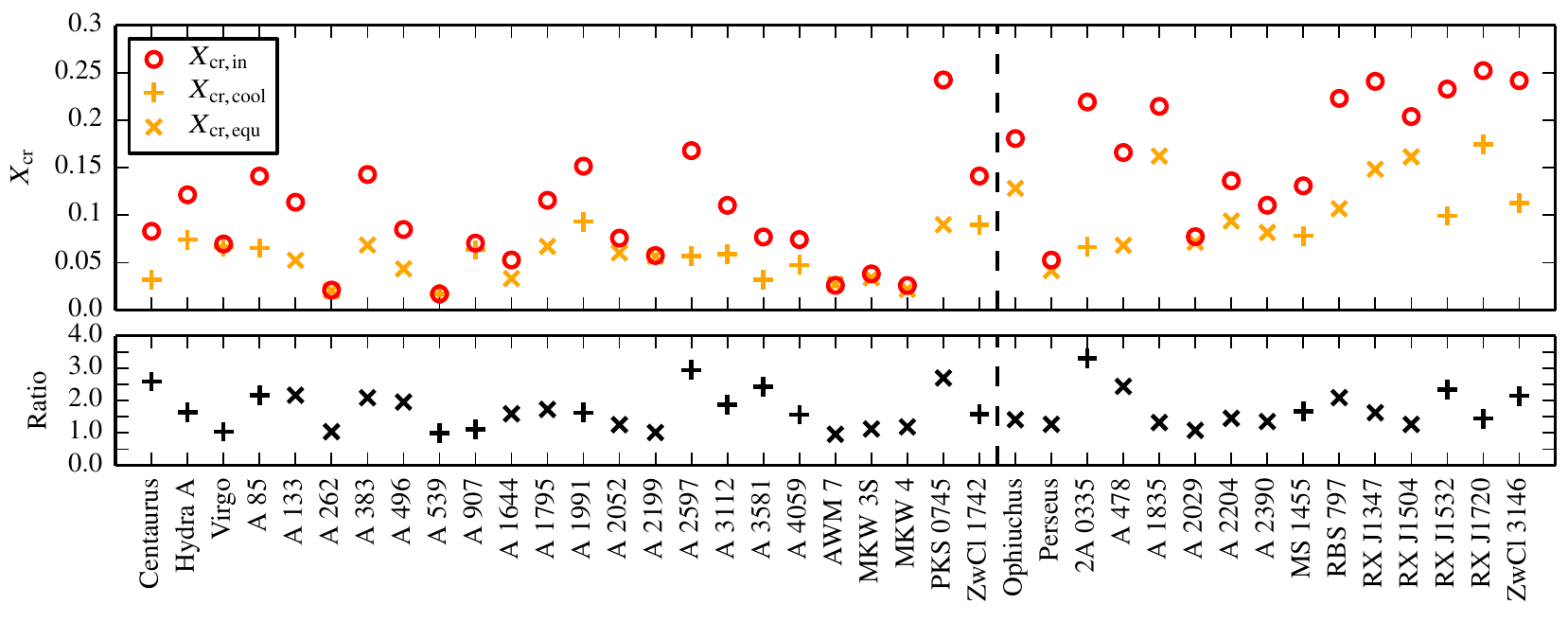}
  \caption{CR-to-thermal pressure ratio $X_{\rmn{cr}}$ at the inner boundary
    (red circle) and at the outer radius of interest, which is either the
    cooling radius (+) or the radius at which the CR and conductive heating
    rates are equal ($\times$). Absolute values for $X_{\rmn{cr}}$ are shown in
    the top panel, the bottom panel shows the ratio of $X_{\rmn{cr}}$ at the
    inner to its value at the outer radius of interest.  The \C{steady state} solutions in
    RMH clusters require larger values of $X_\rmn{cr}$ in comparison to clusters
    without RMHs. $X_\rmn{cr}$ is approximately constant across the radial range
    considered and does not significantly differ in both cluster populations.}
    \label{fig:Xcrcomp}
\end{figure*}

If CR heating balances radiative cooling as in the cluster centres of our steady
state solutions, this imposes further constraints on the non-thermal pressure in
the system. Neglecting the mass accretion rate and thermal conduction (which is
justified in those regions), we get approximately
\begin{equation}
\begin{split}
\mathcal{C}_{\rmn{rad},R} \approx \mathcal{H}_{\rmn{cr},R}
&= -\vv_{\rmn{A},R} \left(\left.\frac{\rmn{d} P_\rmn{cr}}{\rmn{d}r}\right|_R - \frac{5}{2} \frac{\varepsilon_{\rmn{f}} P_{\rmn{cr},R}}{R} \right) \\
& \approx -\vv_{\rmn{A},R} X_\rmn{cr} \left(\left.\frac{\rmn{d} P_\rmn{th}}{\rmn{d}r}\right|_R - \frac{5}{2} \frac{\varepsilon_{\rmn{f}} P_{\rmn{th},R}}{R} \right)
\end{split}
\label{equ:xnth}
\end{equation} 
at a given radius $R$. In the last step, we assume for simplicity that the
CR-to-thermal pressure ratio is constant, which is usually a reasonable
assumption in the inner parts of the cluster.  If temperature and density are
known at the radius $R$, e.g. from observations or from the steady state
solutions, the magnetic field that enters implicitly in the Alfv\'en velocity
and $X_\rmn{cr}$ remain the only unknown quantities. In this case,
Equation~\eqref{equ:xnth} implies $B X_\rmn{cr} = \rmn{const}$ and therefore
$X_B \propto X_\rmn{cr}^{-2}$. We use the values for $X_B$ and $X_\rmn{cr}$ from
the steady state solutions to calculate the constant of proportionality at (i)
the inner boundary of the integration, $r_\rmn{in}$, and (ii) either at the
cooling radius or at the radius where CR and conductive heating are equal. We
choose the smaller of these latter two radii, to avoid that the heating is
dominated by conduction.  The corresponding CR-to-thermal pressure ratio is also
included as $X_{\rmn{cr, r_2}}$ in Table~\ref{tab:params}.

The bottom-right panels of Figs.~\ref{fig:steadystatedemo1} and
\ref{fig:steadystatedemo2} show $X_\rmn{cr}$, $X_B$ and the total
non-thermal-to-thermal pressure ratio $X_\rmn{nt} = X_\rmn{cr}+X_B$ as a
function of $X_\rmn{cr}$. The solid line corresponds to the values at the inner
boundary, $r_\rmn{in}$, and the circle marks the $X_\rmn{cr}$ value that we
obtain for our assumptions of the magnetic field. The dotted line indicates the
non-thermal pressure at the second radius. A plus sign indicates the use of the
cooling radius and the corresponding values of $X_\rmn{cr}$ and $X_B$ (see
Fig.~\ref{fig:steadystatedemo2}). The cross shows that we use the radius at
which CR and conductive heating are equal (see Fig.~\ref{fig:steadystatedemo1}).
Independent of the chosen radius, the lower the magnetic pressure, the higher is
the required CR pressure to realize the balance between heating and cooling and
vice versa. Clearly, the necessary total non-thermal pressure for CR heating to
balance cooling reaches a minimum if the magnetic pressure is half the CR
pressure. In A 1795, the CR-to-thermal pressure ratios that are realized in our
steady state solution are close to this minimum. Hence, the total non-thermal
pressure can not be reduced much further in this cluster. In RX J1720, our
values lie somewhat above the minimum, especially in the centre of the
cluster. However, if the CR pressure is larger than the optimal value, the total
non-thermal pressure only increases linearly with $X_\rmn{cr}$.

\subsection{CR-to-thermal pressure ratio}

How does the CR-to-thermal pressure ratio $X_{\rmn{cr}}$ vary across our sample?
In Fig.~\ref{fig:Xcrcomp}, we show $X_{\rmn{cr}}$ at the inner boundary for each
cluster (red circle) and at the outer radius of interest for CR heating. This is
the smaller radius of either the cooling radius (at which the radiative cooling
time is 1~Gyr) or the radius at which the CR and conductive heating rates are
equal. As already discussed in Section~\ref{sec:steadystatesolutions},
$X_\rmn{cr}$ is approximately constant across the radial range considered and
decreases by at most a factor of 3 towards the outer radius (lower panel of
Fig.~\ref{fig:Xcrcomp}). This behaviour is comparable for clusters with and
without an RMH.

Most interestingly, the upper panel of Fig.~\ref{fig:Xcrcomp} shows that the
CR-to-thermal pressure ratio is typically larger in clusters with an RMH in
comparison to clusters without an RMH with medians of $0.20$ and $0.08$,
respectively. As we will discuss in more detail in \citet{Jacob2016b}, clusters
that host an RMH are on average characterized by higher central densities and
thus a substantially enhanced cooling rate.  To compensate for this increased
cooling in steady state, the CR heating rate and thus $X_\rmn{cr}$ need to be
larger. This is a first indication that the character of the \C{steady state}
solutions is not uniform across our cluster sample and differs for clusters with
and without an RMH.

\C{At first sight, the CR-to-thermal pressure ratios in Fig.~\ref{fig:Xcrcomp}
  appear to be high in comparison to other observational limits on
  $X_{\rmn{cr}}$ that result e.g., from gamma-ray observations of
  clusters. However in our model, the CR source is situated at the cluster
  centre and the CRs lose energy as they stream towards larger radii. This
  implies a steep radial decline of $X_{\rmn{cr}}$ at radii where CR heating is
  insufficient to balance cooling and to maintain the thermal pressure profile
  (equation~\ref{eq:constXcr}). In cluster centres, the CR pressure can only be probed
  in tandem with other non-thermal pressure contributions by comparing
  hydrostatic mass estimates to those inferred by dynamical potential estimates
  that are probed by orbits of stars and globular clusters
  \citep[e.g.,][]{Churazov2010}. These authors conclude that $X_\rmn{cr}$ can
  reach values of 20 -- 30 per cent, which is in agreement with our model.  In
  contrast to that, the upper limits that are derived from the non-detection of
  gamma rays typically assume a global CR population that fills the entire
  cluster out to the virial radius and results from diffusive shock acceleration
  at cosmological formation shocks \citep[e.g.][]{Fermi2014}. Due to the large
  volume that is covered by these models, the allowed $X_\rmn{cr}$ values are
  typically much lower, of the order of 1 -- 2 per cent. We emphasize that in
  order to compare our model with gamma-ray data, we need to compare the
  predicted (radio and gamma-ray) fluxes with the upper limits from
  observations. We carry out such an analysis in our companion paper
  \citep{Jacob2016b}.  }

\subsection{Required fraction of the \textit{Spitzer} conductivity}
\label{sec:condonly}

\begin{figure*}
  \centering
  \includegraphics{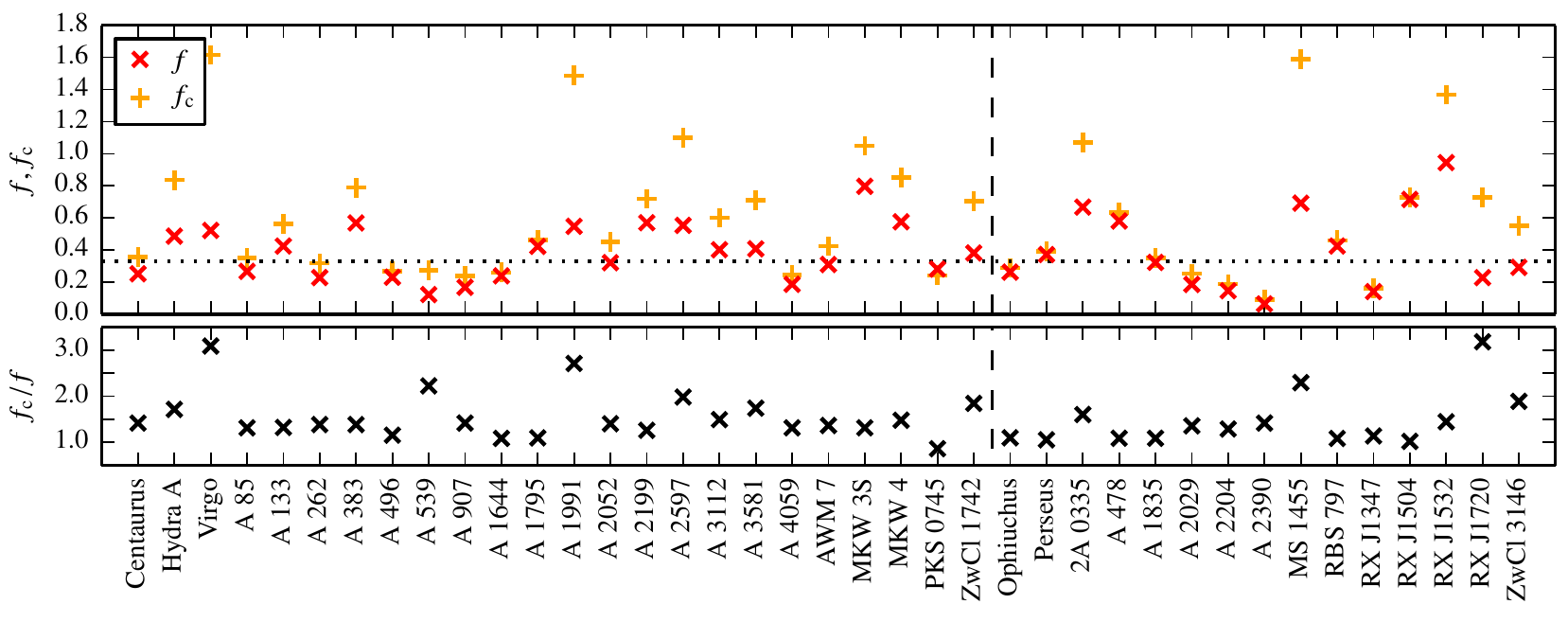}
  \caption{Comparison of the required fraction of the \textit{Spitzer} conductivity in
    the steady state solutions with ($f$) and without ($f_\rmn{c}$) CR
    heating. The top panel shows the absolute values and the bottom panel the
    ratio. If CR heating is added, the required value of $f$ is reduced. For
    some clusters only then $f<1$ can be achieved, but for most clusters CR
    heating has a smaller effect on $f$ since thermal conduction still balances
    radiative cooling on large scales.}
    \label{fig:fvalues}
\end{figure*}

\C{The ICM is a magnetized and weakly collisional medium. It is characterized by
  a mean free path that is many orders of magnitudes larger than the Larmor
  radius. This implies anisotropic transport processes such as thermal
  conduction to act primarily along the orientation of the local magnetic field
  and to dramatically change the way in which the ICM responds to perturbations.
  In the case of a rising temperature profile with radius (which defines cool
  cores), the heat-flux-driven buoyancy instability \citep{Quataert2008} might
  reorient magnetic field lines such that conduction is suppressed in radial
  direction. Instead, hydrodynamic simulations suggest that turbulence,
  e.g. from AGN feedback, is able to randomize the magnetic fields
  \citep{McCourt2011, Yang2016a}.  Moreover, the value of conductivity has been
  estimated on the basis of observations of steep temperature gradients in the
  ICM \citep{Ettori2000} and long-lived cold gas that has been stripped from
  infalling substructure \citep{Eckert2014}. However, these observations can
  also be explained by magnetic draping, which suppresses thermal conduction and
  viscosity across these temperature and density gradients by means of a
  strongly magnetized boundary layer that naturally forms as a result of the
  magneto-hydrodynamics \citep{Lyutikov2006, Dursi2008, Pfrommer2010,
    Ruszkowski2014}.}

\C{The value of the conductivity along the magnetic field is also strongly
  debated.  There is the possibility that mirror instabilities excited by
  pressure anisotropies can act as magnetic traps for the heat-conducting
  electrons, suppressing their transport \citep{Komarov2016,Riquelme2016} or
  that oblique whistler modes can resonate with electrons moving in the
  direction of the heat flux, which potentially causes a suppression of the heat
  flux \citep{RobergClark2016}. However, the effective volume filling fraction
  of these processes has not been studied and it is still unclear whether a
  suppression of the electron transport causes a reduction of the transport of
  thermal energy.}

\C{We treat the fraction of the \textit{Spitzer} conductivity as an
  eigenvalue of the system of steady state equations. Red crosses in the top panel of Fig.~\ref{fig:fvalues} show the \textit{Spitzer}
  fractions that we obtain from our fiducial solutions. Most of the values lie
  between $0.2$ and $0.6$. Note that we directly exclude solutions with $f > 1$
  as being unphysical. Still, our values are somewhat on the high side as
  indicated by the dashed line that represents the isotropic average $f =
  0.33$. Nevertheless, considering the ongoing debate about the conductivity in
  the ICM, so far there is no major problem with our results for $f$. This
  result is in line with findings by \citet{Voit2015} who suggest that thermal
  conduction appears to be important for distinguishing clusters with and
  without a cool core.}

To analyse the impact of CR heating, we also solve the system of hydrodynamic
equations without CRs, i.e.  Equations~\eqref{equ:contss}, \eqref{equ:eulerss},
\eqref{equ:gasenss} and \eqref{equ:fourierss}. Radiative cooling is then
balanced only by thermal conduction, which was already explored by
\citet{Zakamska2003} and \citet{Guo2008globalstability}.  Unlike these authors,
we supplement the gravitational potential of the NFW profile by that of an SIS
at small radii, which appears to be required by dynamical potential estimates
\citep{Churazov2010}. We use parameters for the gravitational potential
(i.e. $M_\rmn{s}$, $r_\rmn{s}$, $r_\rmn{t}$), the mass accretion rate, the
radial range and the boundary conditions that we describe in
Section~\ref{sec:steadystateequ}. Without CRs, the temperature and density
profiles do not change much since they are primarily determined by the
gravitational potential and the boundary conditions.

Interestingly, we obtain different fractions of the \textit{Spitzer} conductivity,
denoted as $f_\rmn{c}$, which are shown in Fig.~\ref{fig:fvalues} as orange plus
signs.  As expected, the required conductivity increases in comparison to the
results with CRs since now conduction alone has to balance cooling; in some
cases resulting in a conductivity significantly exceeding the \textit{Spitzer}
value. Thus, CR heating is required in those clusters to achieve a fraction of
the \textit{Spitzer} conductivity that is smaller than unity.  The bottom panel in
Fig.~\ref{fig:fvalues} shows the ratio of the \textit{Spitzer} conductivity with and
without CRs. It can be seen that for many clusters the fraction of the \textit{Spitzer}
conductivity is not altered dramatically by the addition of CRs, only in rare
cases more than a factor of two. The reason is that in both models thermal
conduction balances radiative cooling on large scales in many clusters. An
example is A~1795, in which the required conductivity remains almost the
same. However, if CR heating dominates also on larger scales in our fiducial
model, as in our second example cluster RX~J1720, the required fraction of the
\textit{Spitzer} conductivity is significantly reduced in comparison to the
conduction-only case.

Despite the seemingly small effects of CR heating on the steady state solutions,
our fiducial model with CR heating has some clear advantages over the model that
only includes thermal conduction. First, CR heating is locally stable at
certain temperatures in contrast to thermal conduction
\citep{Pfrommer2013}. Moreover, CR heating enables a self-regulated AGN feedback
loop: CRs are injected by the central AGN and heat the cluster gas by streaming
outwards. As soon as the CR population is too dilute and has lost most of its
energy, radiative cooling overcomes CR heating such that cold gas can fuel the
AGN and trigger CR injection again.


\subsection{Parameters for maximal CR heating}
\label{sec:parameters}

The model parameters that enter the steady state equations before integration
have a large impact on the solution. Thus, we scrutinize our choice of
parameters in this section. To this end, we distinguish between clusters with and
without RMHs. Here, we analyse correlations between our parameters and observed
quantities and discuss the amount of fine-tuning in our solutions.

We model the gravitational potential as a superposition of a singular isothermal
sphere in the centre of the cluster and an NFW profile at larger radii
(see also \C{Fig.~\ref{fig:gravpot} in} Section~\ref{sec:modelspec}). The radius
$r_\rmn{t}$ determines the transition between these potentials \C{as it
  delineates -- by construction -- equal force contributions by the
  gravitational potentials of the SIS and the NFW profiles}. To justify the
usage of an isothermal sphere within the transition radius, we compute the
temperature difference between the temperature at the selected transition radius
and the inner radius for each cluster. We find that $\Delta T (r_\rmn{in},
r_\rmn{t}) \le 1~\rmn{keV}$ except for one cluster with a temperature difference
of $1.7~\rmn{keV}$. Assuming quasi-hydrostatic equilibrium, this demonstrates
that within $r_\rmn{t}$ an isothermal sphere is a valid assumption for all
clusters.

\subsubsection{Parameter correlations}
\label{sec:correlations}

\begin{figure*}
  \centering
  \includegraphics{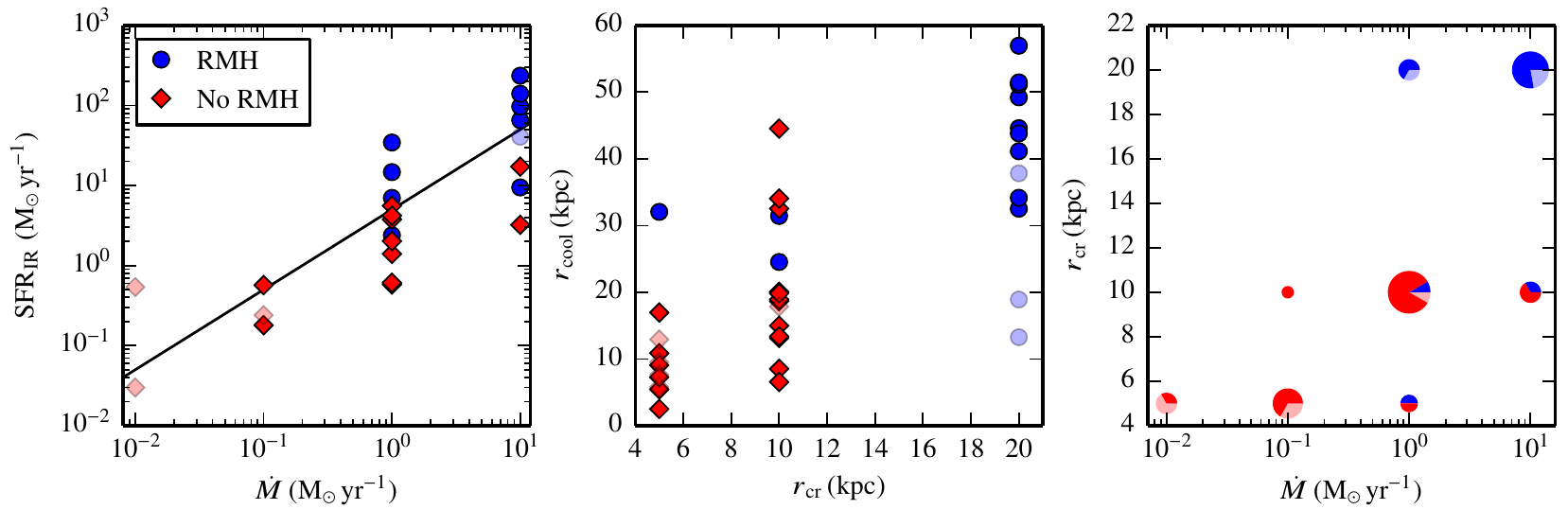}
  \caption{Correlation between model parameters and observed
    quantities. \textit{Left.} Comparison between observed infra-red SFR and
    mass accretion rate of our steady state solution. Larger SFRs imply larger
    mass accretion rates and are more likely to be accompanied by the phenomenon
    of RMHs. The black line indicates a linear relation between SFRs and
    $\dot{M}$.  \textit{Middle.} We compare the cooling radius $r_\rmn{cool}$ to
    the CR injection radius $r_\rmn{cr}$. The larger the cooling radius, the
    larger is the selected CR injection radius, albeit with substantial
    scatter. Clusters with RMHs have the largest cooling and CR injection radii.
    \textit{Right.} As a consequence of the other relations, the CR injection
    radii and mass accretion rates are also correlated.}
    \label{fig:MdotvsSFR}
\end{figure*}

In the discussion of our cluster sample, we already pointed out a correlation
between the cooling radius and the observed infra-red SFR in
Section~\ref{sec:coolingtime}.  Here, we pursue this topic further and show the
most interesting relations between the model parameters as well as between
parameters and observations in Fig.~\ref{fig:MdotvsSFR}.

In the left-hand panel of Fig.~\ref{fig:MdotvsSFR}, we compare the observed infra-red
SFRs as listed in Table~\ref{tab:sample} with the mass accretion rates $\dot{M}$
from our steady state solutions. Clusters with high SFRs require higher mass
accretion rates to obtain a steady state solution in which CR heating
dominates. For visual guidance, the black line shows a linear relation between
both quantities, which is consistent with the data.  Apparently, larger mass
deposition rates are able to sustain larger SFRs. However, star formation and the
mass accretion considered here operate on very different time and length scales,
such that a direct link between both is not necessarily expected. Moreover, we
note that the SFRs are roughly a factor of 10 higher than the mass accretion
rates. Partly this may result from our upper limit of the accretion rate of
$\dot{M} = 10\,\rmn{M_{\sun}\,yr^{-1}}$. \C{Another possibility is that star formation is triggered by the interaction of AGN jets and the ambient medium \citep{Gaspari2012a, Li2014, Brighenti2015}.} Interestingly, clusters with an RMH have
higher observed SFRs and therefore higher values of $\dot{M}$ in comparison to
clusters without RMHs.

The middle panel of Fig.~\ref{fig:MdotvsSFR} shows the relation between the
cooling radius $r_\rmn{cool}$ as defined in Section~\ref{sec:coolingtime} and
the radius $r_\rmn{cr}$ in the CR source function. The larger the cooling
radius, the larger is the required value of $r_\rmn{cr}$. Such a relation can be
expected since a large cooling radius implies that the region where the cooling
problem is most severe is also large. In order for CR heating to stably balance
radiative cooling, comparably large amounts of CRs are needed in this entire
region, which calls for a large CR injection radius $r_\rmn{cr}$.  As before,
there are differences between the populations of galaxy clusters with and
without RMHs: both radii are significantly larger for clusters hosting RMHs.

We finally show the CR injection radius $r_\rmn{cr}$ as a function of the mass
accretion rate $\dot{M}$ in the right-hand panel in Fig.~\ref{fig:MdotvsSFR}. Since
both values are discrete in the parameter grid, points often lie on top of each
other. Thus, we indicate the number of the overlying points with the area of the
pie chart and the colours show the contributions from clusters with and without
RMHs. The smallest pie charts contain only one cluster, whereas the largest chart
represents 12 clusters. For the majority of clusters the mass accretion rate
correlates well with the CR injection radius, which is a consequence of the
relations presented before: larger mass accretion rates imply larger SFRs, which
in turn imply larger cooling radii (see Fig.~\ref{fig:rcoolvsSFR}) and hence
also larger CR injection radii. As expected, clusters hosting an RMH are
characterized by higher mass accretion rates and larger CR injection radii in
comparison to cluster without an RMH.

These correlations are reassuring in that our parameter choices reflect the
observed trends and relations of SFR and cooling radius of
Fig.~\ref{fig:rcoolvsSFR}.  Moreover, the relations demonstrate that there is
some diversity in the population of CC clusters, with a continuous sequence from
cluster hosting an RMH to those without: the latter population shows a smaller
SFR and mass accretion rate, which may indicate that CR heating is more
efficient in those clusters. Although the smallest mass accretion rates occur in
clusters with very low masses, all trends are still clearly visible for the
subsample of clusters with similar masses.

\subsubsection{Discussion of fine-tuning}

The existence of global solutions immediately poses the question of potential
fine-tuning of the parameter values. Given the complexity of the involved
physics that includes modelling the gravitational potential, the magnetic field
and the CR population, a comparably large number of parameters is
unavoidable. For some parameters we use values that are common to all
clusters. However, we put down a four-dimensional parameter grid in which we are
searching for viable solutions.

Due to the diversity in various cluster parameters, such as cooling radius,
SFR, and also to a certain extent mass (see Figs.~\ref{fig:M200Redshift} and
\ref{fig:rcoolvsSFR}), we do not expected that all their properties can be
described by the same universal parameters. Comparable observational constraints
among our cluster sample would help to substantiate the parameter
choices. However, observations of the magnetic fields or CR populations for such
a large cluster sample are not feasible in the near future such that a
parametrization of these quantities remains necessary.

Our parameter grid can span orders of magnitude in a parameter value (especially
for $\dot{M}$ and $\varepsilon_\rmn{cr}$). Hence, our parameter choices
represent a range of parameters rather than fine-tuned precise values and we get
similarly well matches to the observational profile if we vary our best-fitting
values somewhat. This leads us to the conclusion that although some tuning is
indeed required for our solutions, extensive fine-tuning is not necessary.

\section{Summary and conclusions}
\label{sec:conclusions}

The cooling flow problem in CC clusters remains one of the most
interesting open questions in galaxy clusters. While the paradigm of
self-regulated AGN feedback is very attractive, the physical heating mechanism
that balances radiative cooling has not yet been identified. In this work, we
have analysed whether a combination of CR heating and thermal conduction is able
to provide the required heating.

To this end, we have compiled one of the largest samples of CC clusters ever
used for a theoretical investigation of the cooling flow problem. Here, we have
focused on clusters for which non-thermal activity has either already been
observed or which are predicted to be suitable targets for non-thermal emission.
In particular, this includes all clusters that host a radio mini halo, i.e., an
extended radio emission in the centres of the clusters.  Clusters with an RMH
are typically at slightly higher redshifts than clusters without RMHs, but the
virial masses of most clusters are comparable with some outliers that we treat
separately.  We find that the observed infra-red SFR and the cooling radius,
which we define as the radius where the cooling time equals $1~\rmn{Gyr}$, are
correlated. Moreover, clusters with an RMH have larger SFRs and cooling radii
than clusters without RMHs.

For all clusters, we found steady state solutions to the system of hydrodynamic
equations coupled to the CR energy equation. The thermal energy equation
accounts for thermal conduction as well as Alfv\'en wave heating excited by
streaming CRs. We choose the parameters of the gravitational potential, CR
streaming and injection to obtain physical solutions and ask for maximum CR
heating solutions. In consequence, we find solutions that match the observed
density and temperature profiles well\C{, however requiring a somewhat high
  conductivity for some systems}. Radiative cooling is typically balanced by CR
heating in the cluster centres and by thermal conduction in the intermediate
cluster parts, closer to the peak in temperature. The combination of these two
heating mechanisms has several advantages over models that include only one of
the two processes. CR heating is locally stable \C{at temperature values
  corresponding to islands of stability that form at locations of cooling line
  complexes in the cooling function \citep{Pfrommer2013}} and it allows for
self-regulated AGN feedback, in contrast to thermal conduction, which appears to
be nonetheless required to balance cooling at large scales and to allow for mass
deposition rates that are in agreement with observational findings.

Our solutions predict modest mass deposition rates; consistent with the low star
formation rates and the observed reservoirs of cold gas in the centres of those
systems. The cooling gas can escape the detection of soft X-rays
($kT\lesssim0.5$~keV) by absorption in the filaments with a sufficiently high
integrated hydrogen column density and/or by mixing the cooling gas with colder
gas, thereby lowering its temperature non-radiatively \citep{Werner2013,
  Werner2014}.

Furthermore, we used our comparably large cluster sample to analyse the
parameters of these steady state solutions. We found weak correlations between
the observed infra-red SFR and the mass deposition rate in our solutions as well
as between the cooling radius and the radial extent of the CR
injection. Particularly, clusters with and without RMHs occupy different parts
of these relations.  Clusters that are hosting an RMH have higher star formation
and mass accretion rates in comparison to clusters without an RMH. In addition,
the cooling and CR injection radii are typically larger in clusters with an RMH.
Hence, the existence of an RMH delineates a homogeneous subclass within the
population of CC clusters.

In this work, we present steady state solutions that match X-ray observations
well. However, these solutions predict a CR population that interacts
hadronically with the ambient medium.  As a result, pions are produced which
decay into electrons and photons that can be observed in the radio and
gamma-ray regime, respectively. The crucial question whether the CR
populations of our solutions are in agreement with current observations and
upper limits of this non-thermal emission will be addressed in our companion
paper \citep{Jacob2016b}.

\section*{Acknowledgements}

We thank Volker Springel for his helpful comments, Norbert Werner for providing
the data for the Ophiuchus cluster\C{, and an anonymous referee for a
  constructive report}.  SJ acknowledges funding through the graduate college
\textit{Astrophysics of cosmological probes of gravity} by
Landesgraduiertenakademie Baden-W\"urttemberg. CP acknowledges support by the
ERC-CoG grant CRAGSMAN-646955. Both authors have been supported by the Klaus
Tschira Foundation.


\bibliographystyle{mnras}
\bibliography{bib}


\appendix

\section{Additional tables and figures}
\label{sec:app}

\begin{figure*}
	\centering
	\includegraphics{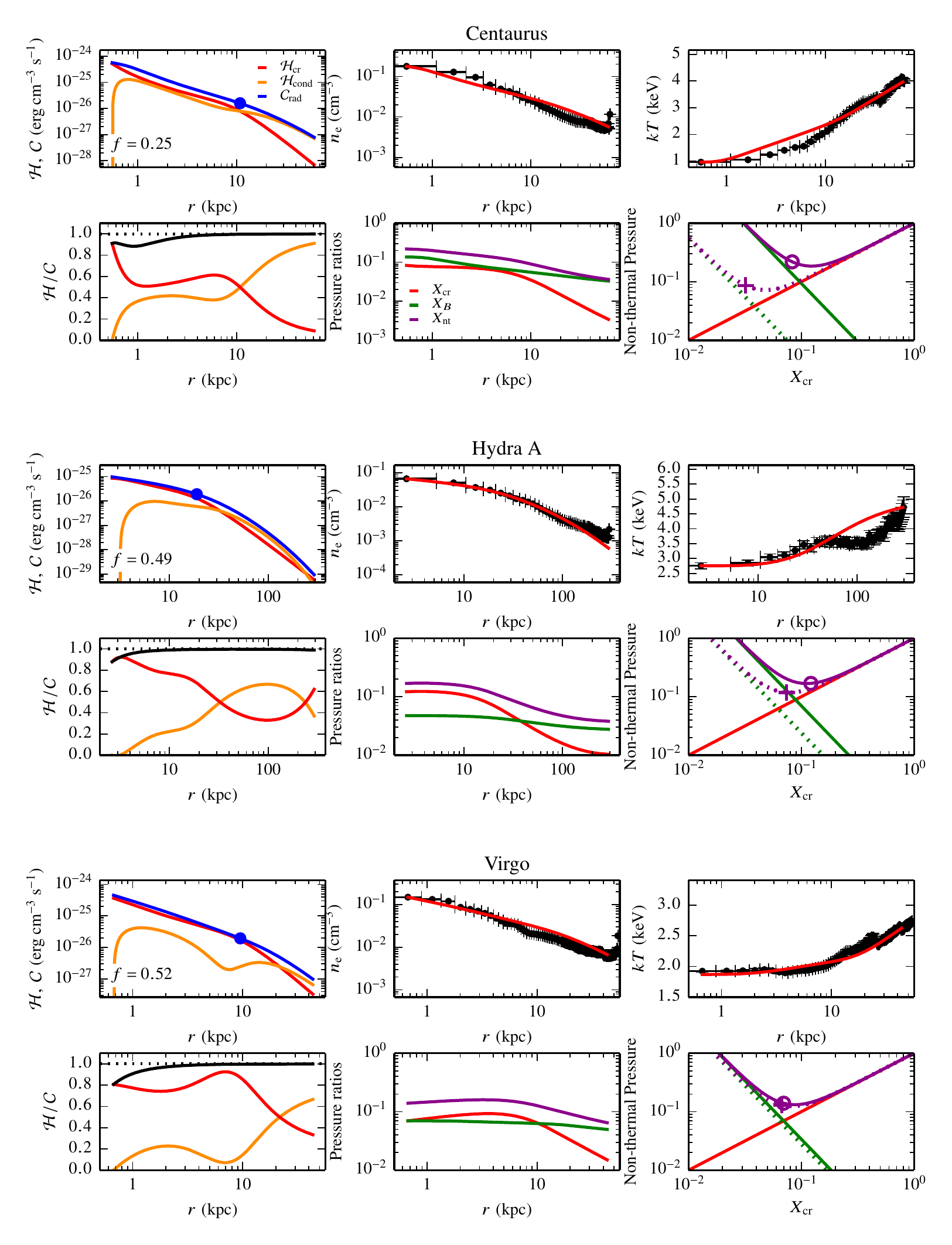}
    	\caption{We show the same properties of the steady state solutions as in Fig.~\ref{fig:steadystatedemo1} for different clusters.}
    \label{fig:steadystateall1}
\end{figure*}

\begin{figure*}
	\centering
	\includegraphics{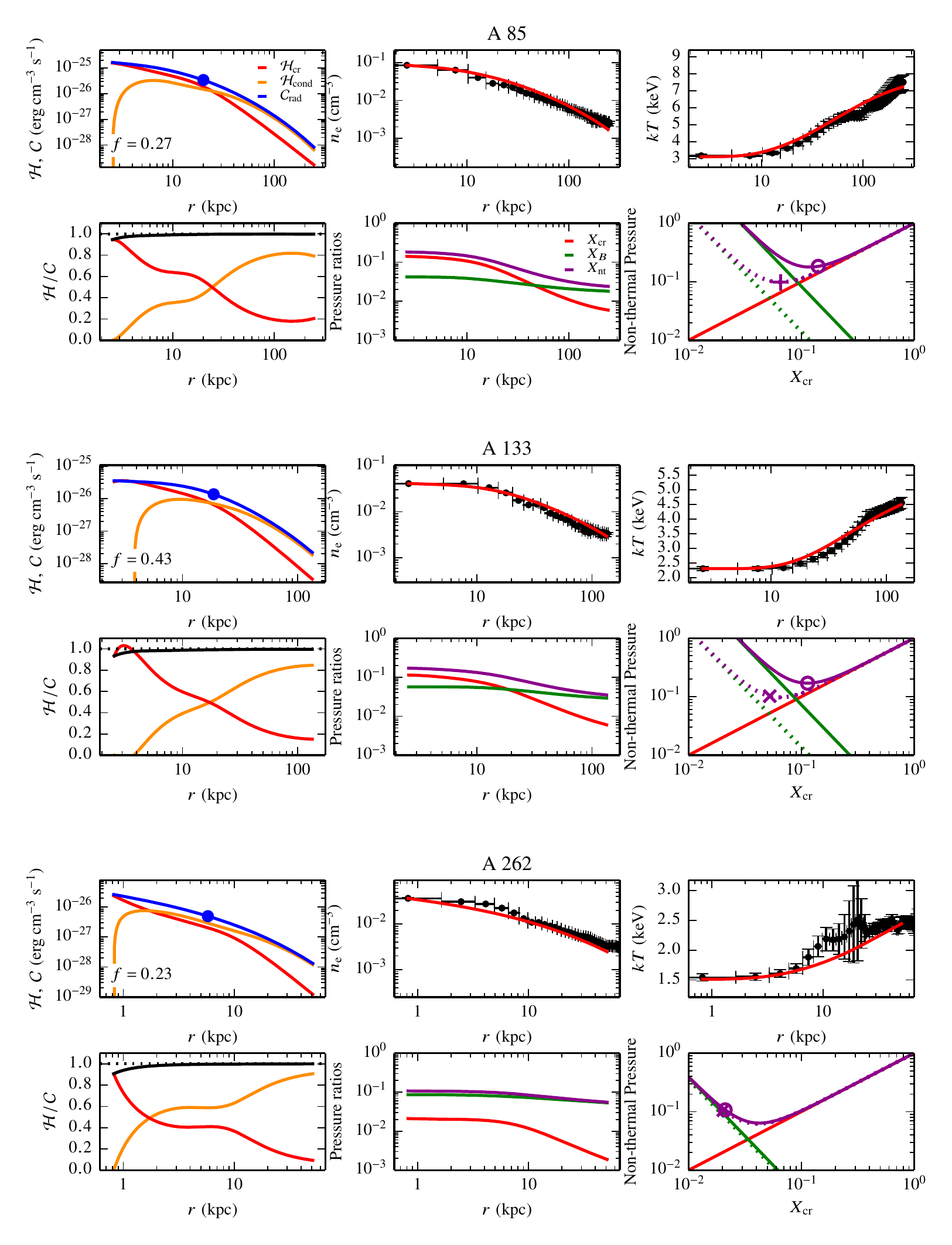}
    	\caption{We show the same properties of the steady state solutions as in Fig.~\ref{fig:steadystatedemo1} for different clusters.}
    \label{fig:steadystateall2}
\end{figure*}

\begin{figure*}
	\centering
	\includegraphics{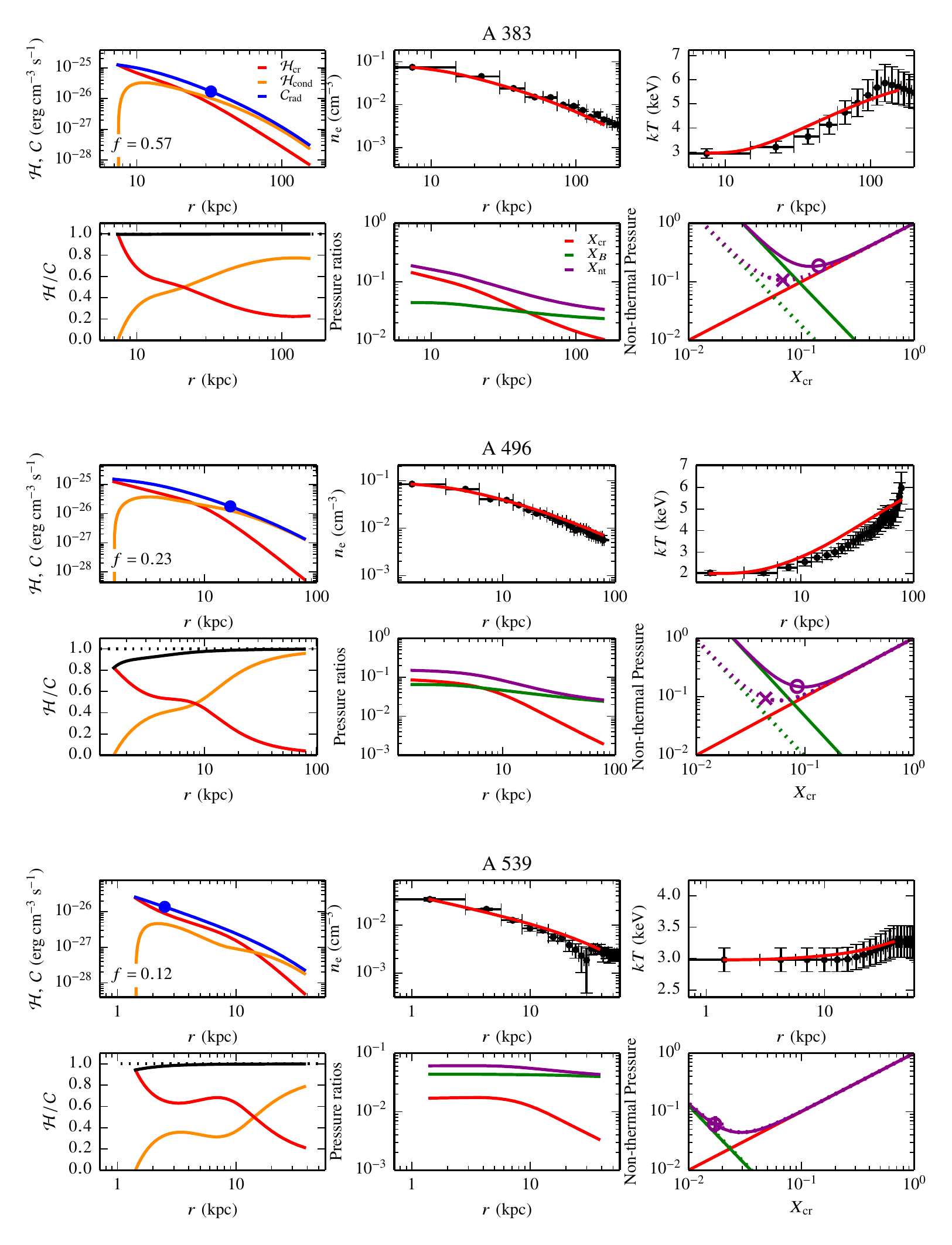}
    	\caption{We show the same properties of the steady state solutions as in Fig.~\ref{fig:steadystatedemo1} for different clusters.}
    \label{fig:steadystateall3}
\end{figure*}

\begin{figure*}
	\centering
	\includegraphics{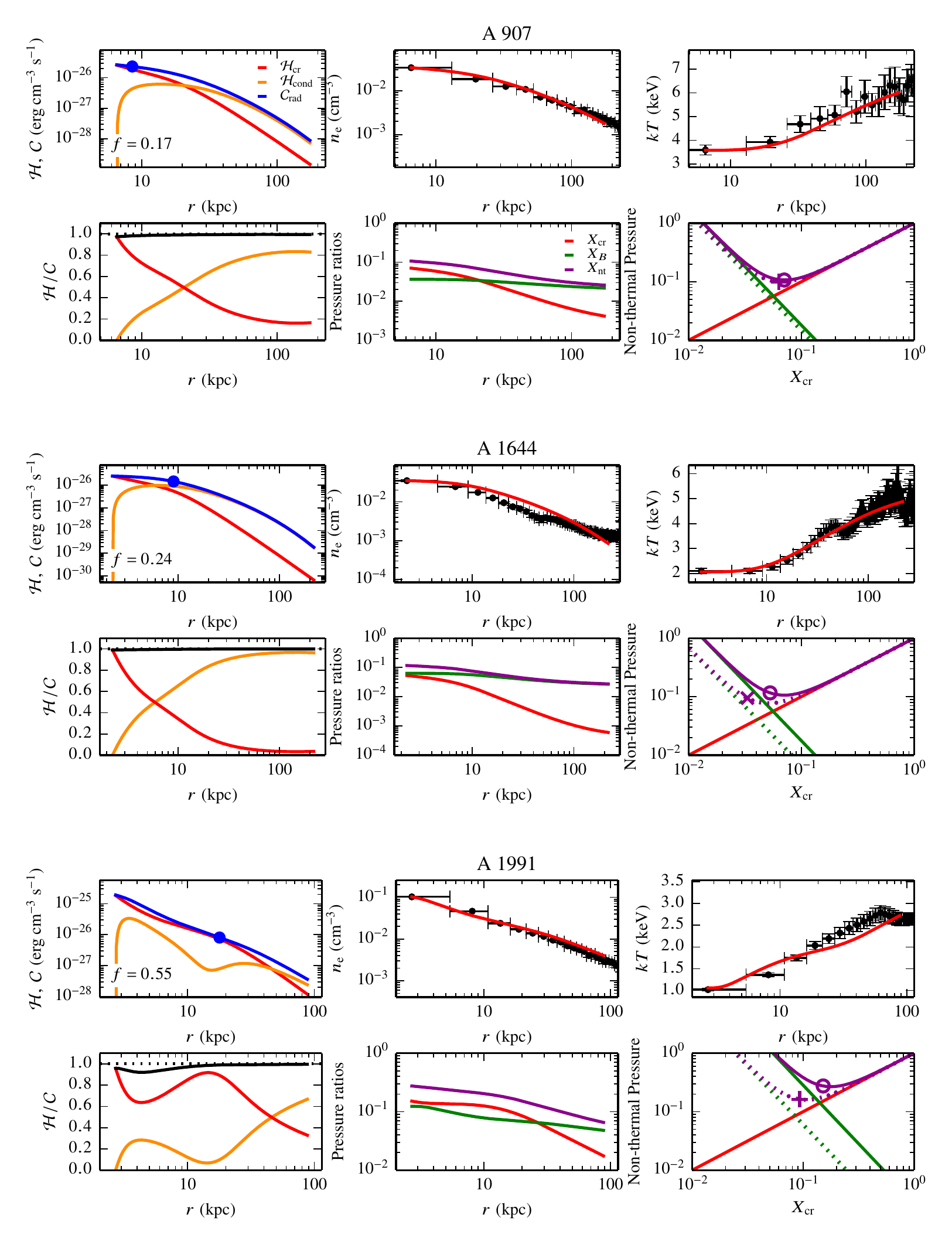}
    	\caption{We show the same properties of the steady state solutions as in Fig.~\ref{fig:steadystatedemo1} for different clusters.}
    \label{fig:steadystateall4}
\end{figure*}

\begin{figure*}
	\centering
	\includegraphics{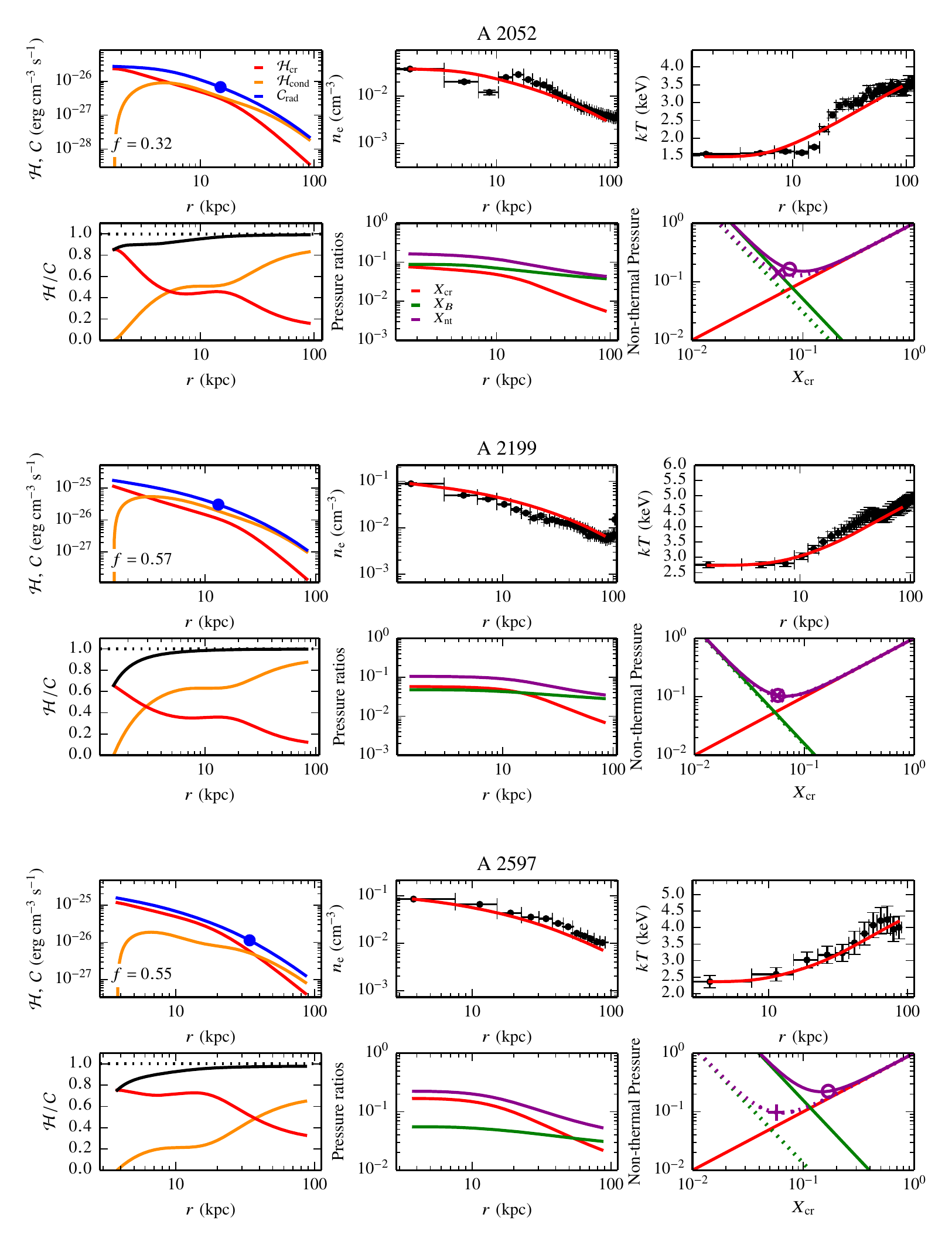}
    	\caption{We show the same properties of the steady state solutions as in Fig.~\ref{fig:steadystatedemo1} for different clusters.}
    \label{fig:steadystateall5}
\end{figure*}

\begin{figure*}
	\centering
	\includegraphics{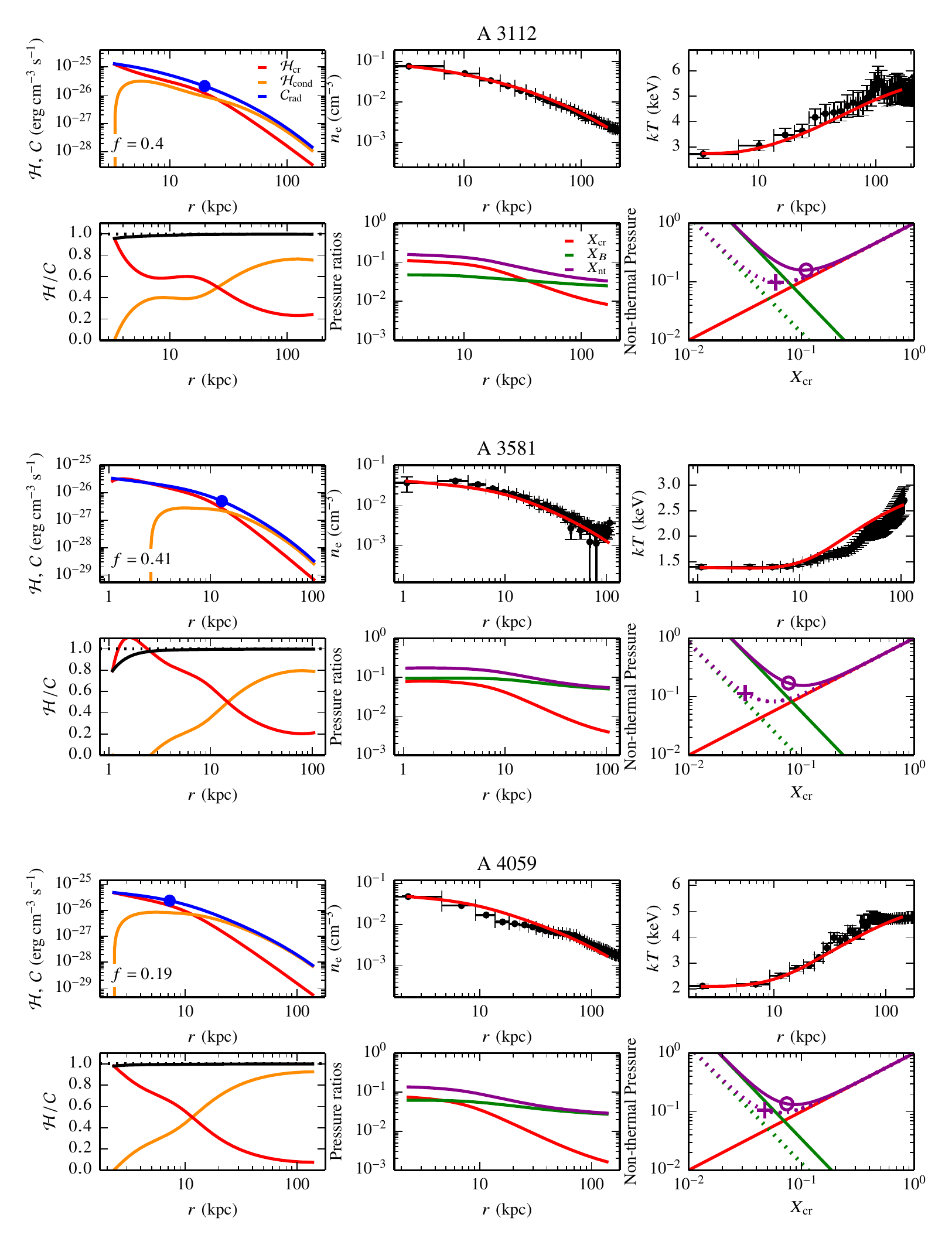}
    	\caption{We show the same properties of the steady state solutions as in Fig.~\ref{fig:steadystatedemo1} for different clusters.}
    \label{fig:steadystateall6}
\end{figure*}

\begin{figure*}
	\centering
	\includegraphics{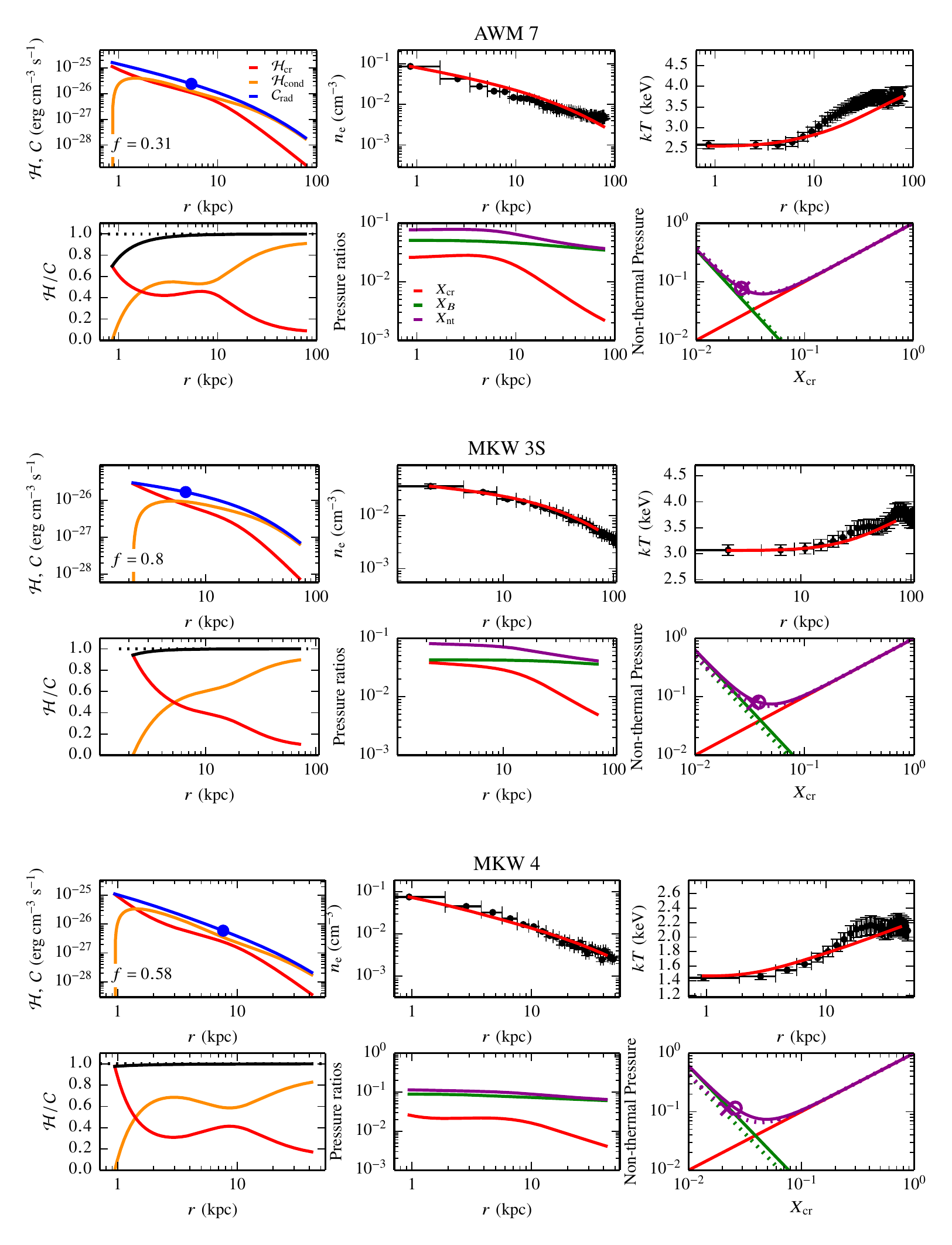}
    	\caption{We show the same properties of the steady state solutions as in Fig.~\ref{fig:steadystatedemo1} for different clusters.}
    \label{fig:steadystateall7}
\end{figure*}

\begin{figure*}
	\centering
	\includegraphics{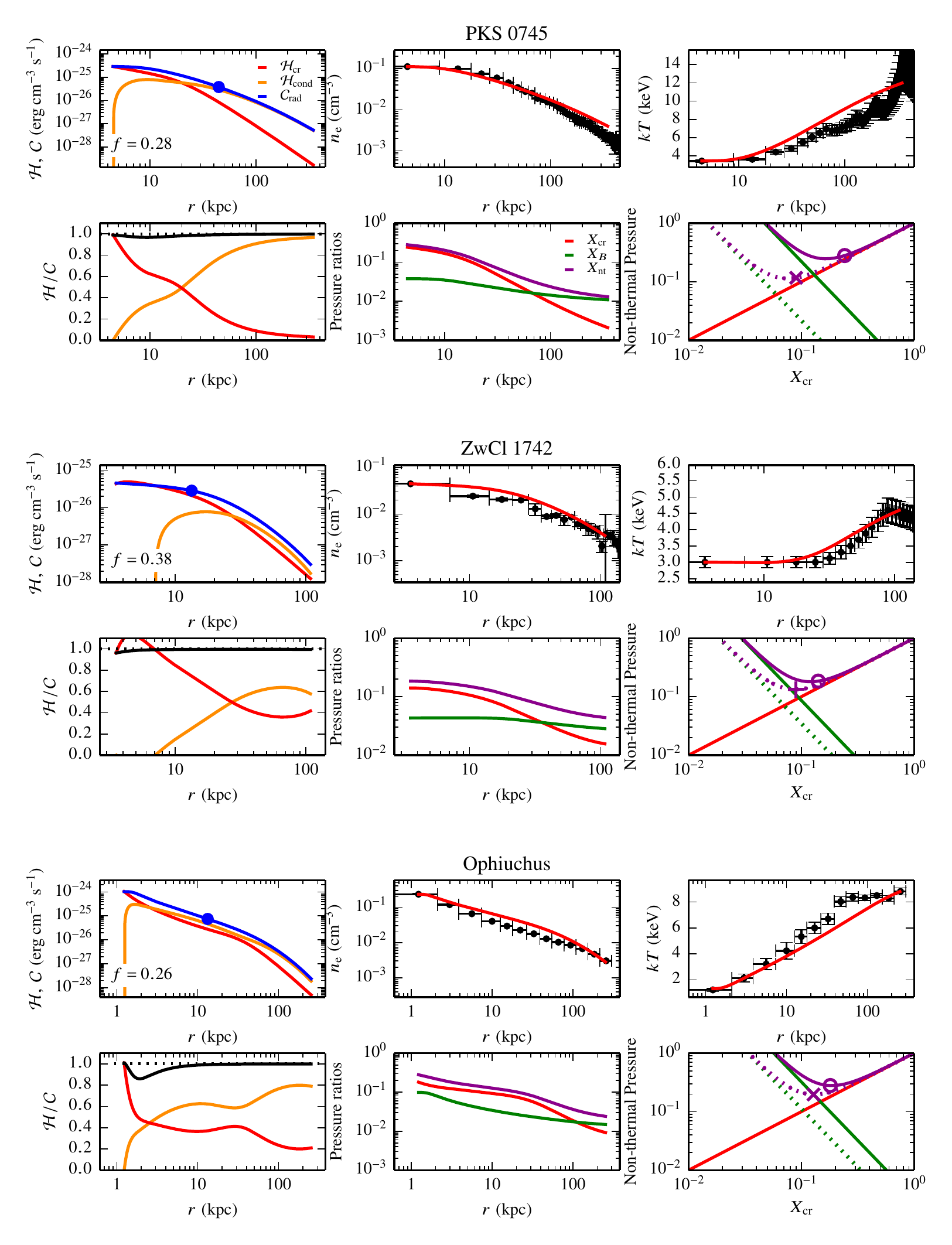}
    	\caption{We show the same properties of the steady state solutions as in Fig.~\ref{fig:steadystatedemo1} for different clusters.}
    \label{fig:steadystateall8}
\end{figure*}

\begin{figure*}
	\centering
	\includegraphics{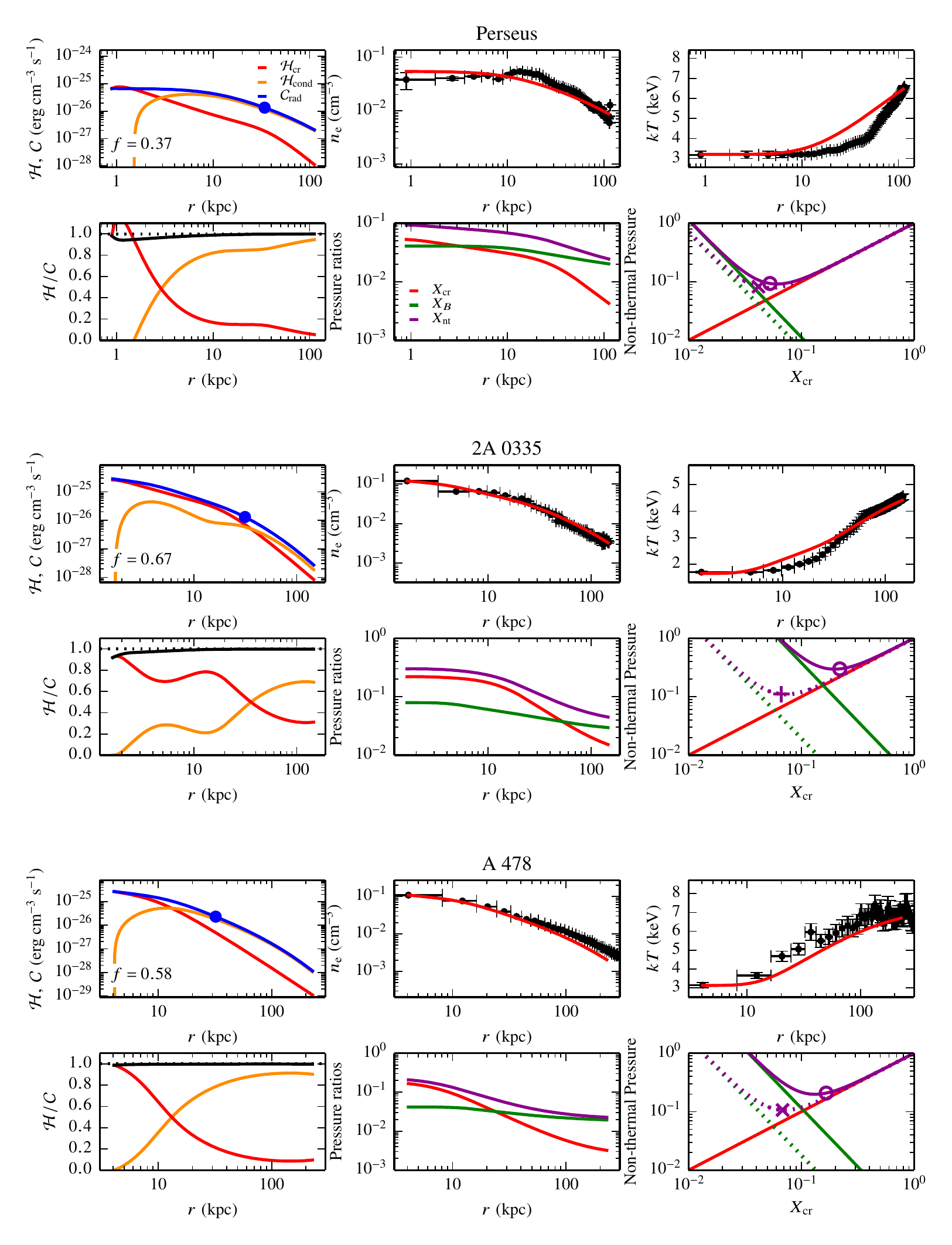}
    	\caption{We show the same properties of the steady state solutions as in Fig.~\ref{fig:steadystatedemo1} for different clusters.}
    \label{fig:steadystateall9}
\end{figure*}

\begin{figure*}
	\centering
	\includegraphics{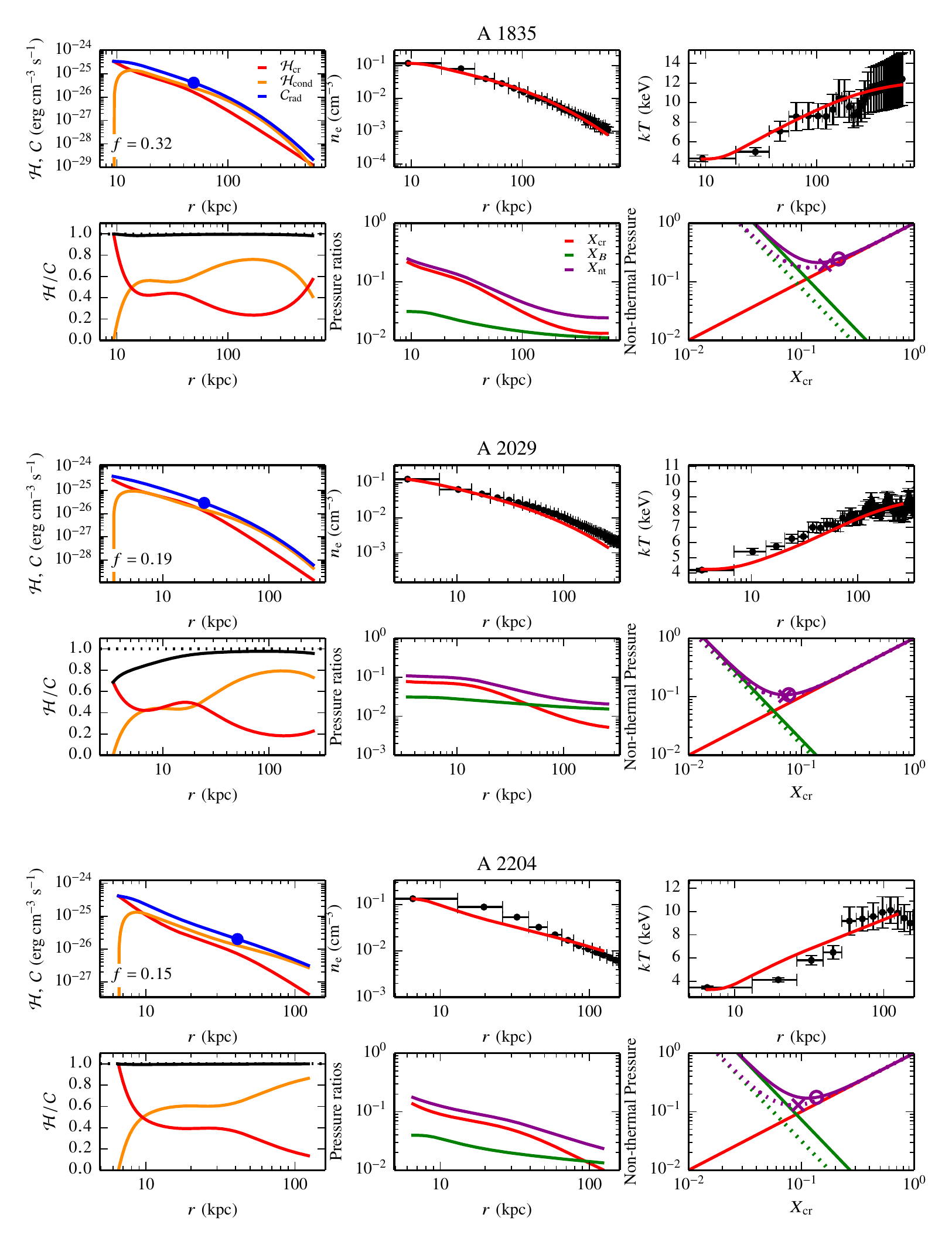}
    	\caption{We show the same properties of the steady state solutions as in Fig.~\ref{fig:steadystatedemo1} for different clusters.}
    \label{fig:steadystateall10}
\end{figure*}

\begin{figure*}
	\centering
	\includegraphics{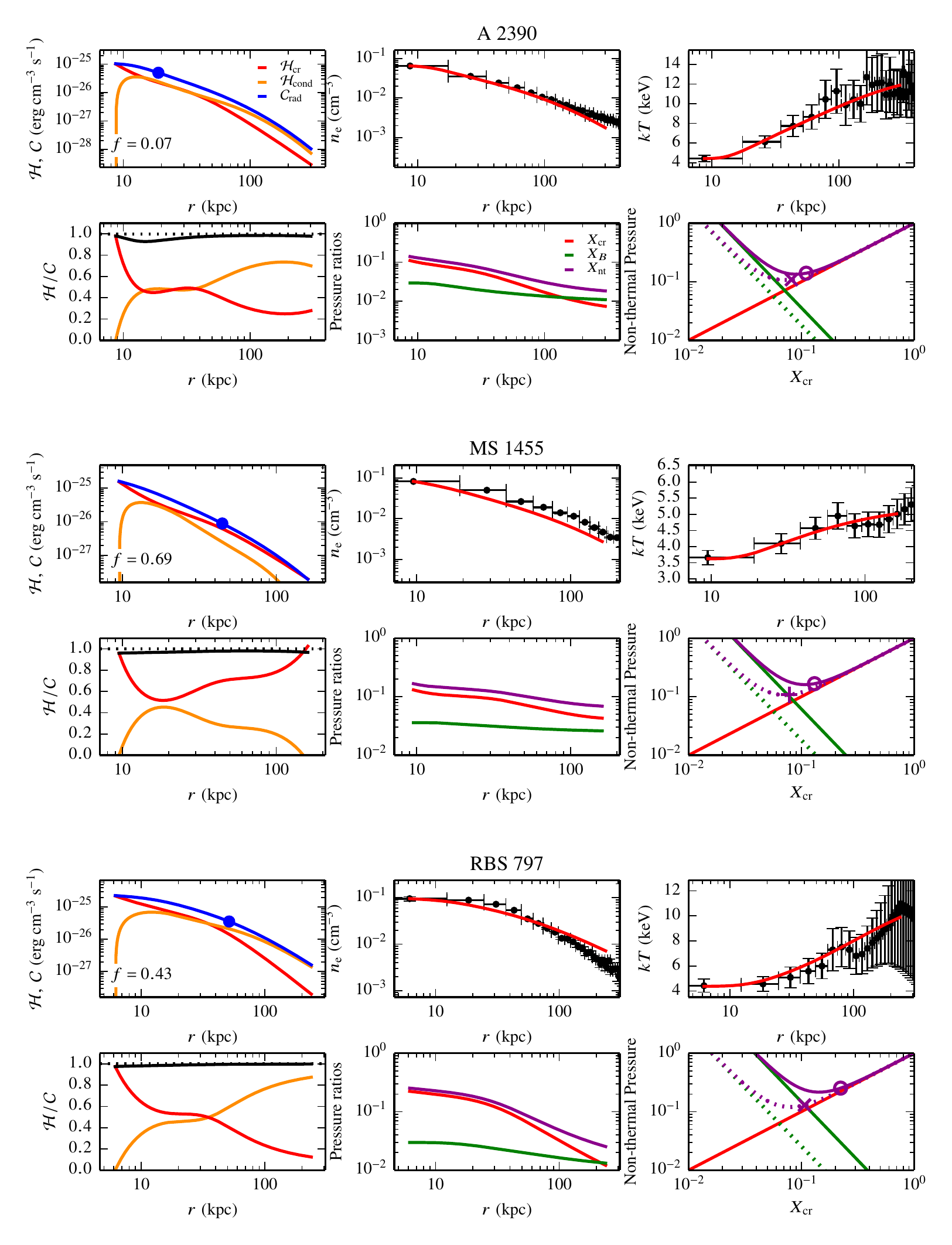}
    	\caption{We show the same properties of the steady state solutions as in Fig.~\ref{fig:steadystatedemo1} for different clusters.}
    \label{fig:steadystateall11}
\end{figure*}

\begin{figure*}
	\centering
	\includegraphics{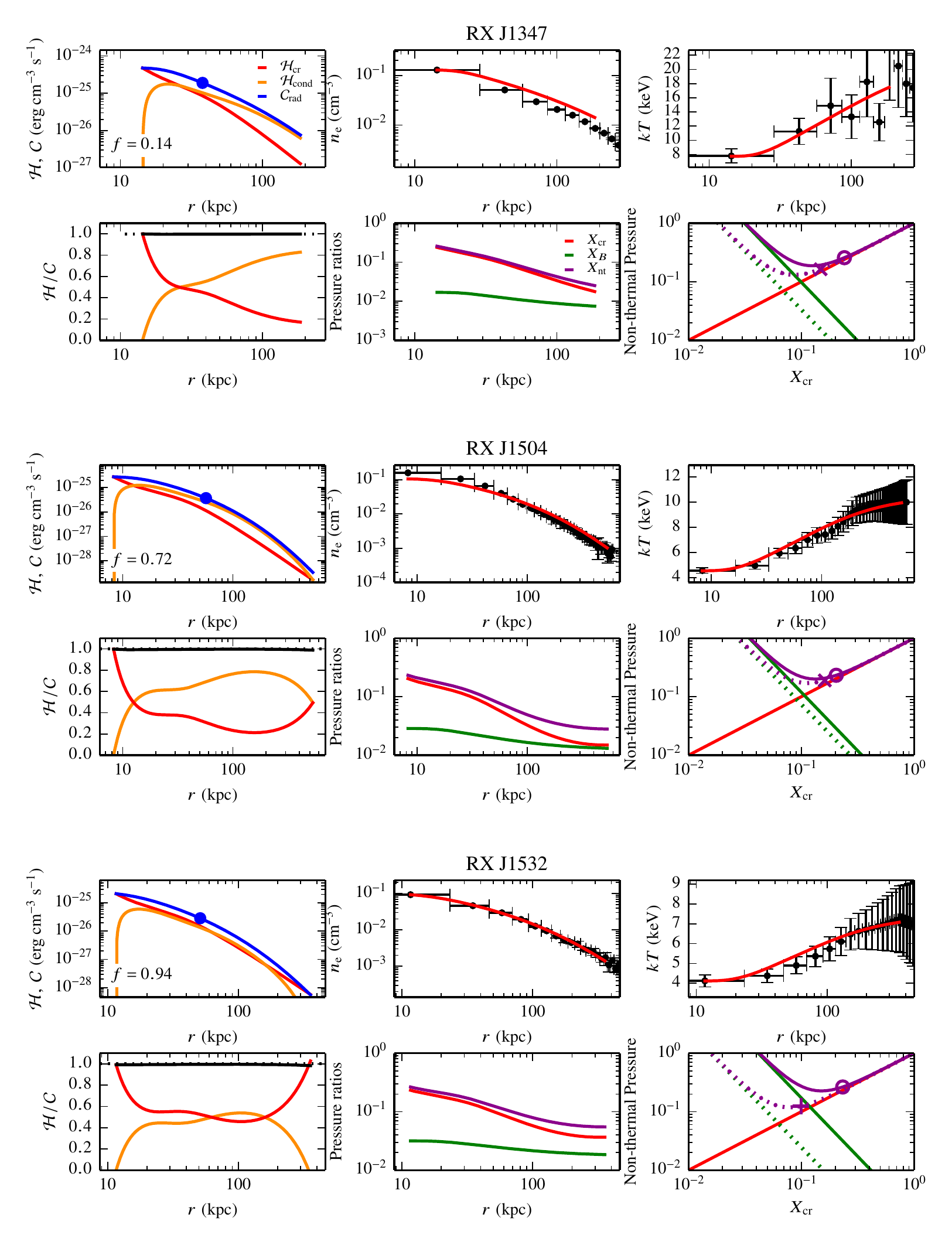}
    	\caption{We show the same properties of the steady state solutions as in Fig.~\ref{fig:steadystatedemo1} for different clusters.}
    \label{fig:steadystateall12}
\end{figure*}

\begin{figure*}
	\centering
	\includegraphics{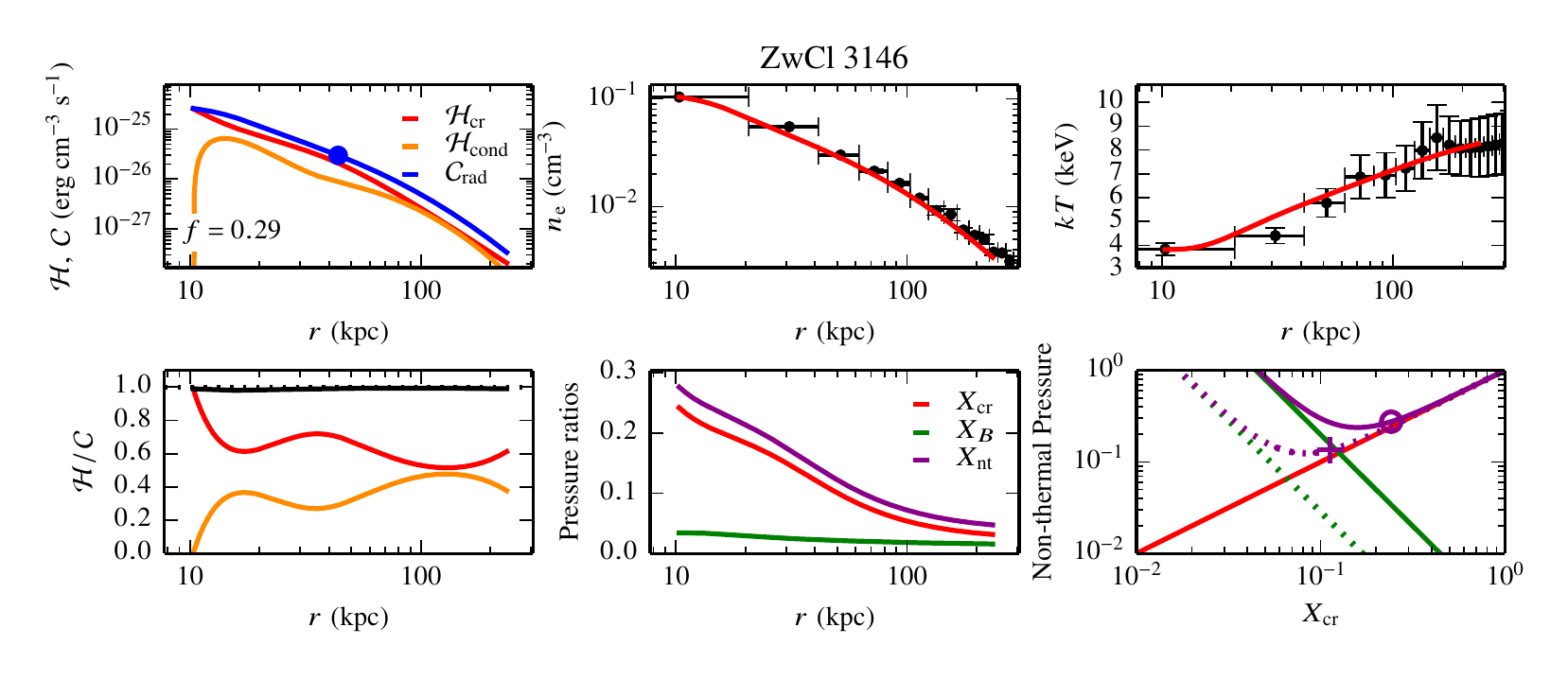}
    	\caption{We show the same properties of the steady state solutions as in Fig.~\ref{fig:steadystatedemo1} for different clusters.}
    \label{fig:steadystateall13}
\end{figure*}

We show the fit parameters for the temperature profile in Table~\ref{tab:fitparam}. Figs.~\ref{fig:steadystateall1} - \ref{fig:steadystateall13} show various aspects of the steady state solutions by analogy to Figs.~\ref{fig:steadystatedemo1} and \ref{fig:steadystatedemo2} (albeit in a different order) for the remaining clusters in our sample. The clusters are ordered as in Table~\ref{tab:sample}. The density and temperature data for Ophiuchus are weighted averages of the sector profiles provided by \citet{Werner2016}.

\begin{table}
\caption{Parameters for temperature profiles.}
\label{tab:fitparam}
\begin{threeparttable}
\begin{tabular}{l  r r r r r r r}
\hline
	 Cluster			&\multicolumn{1}{c}{$r_{\mathrm{cut},T}^{\rmn{(1)}} $} 	&\multicolumn{1}{c}{$a^{\rmn{(2)}}$}	&\multicolumn{1}{c}{$T_{\rmn{0}}$}	&\multicolumn{1}{c}{$ T_{\rmn{1}}$} &\multicolumn{1}{c}{$r_{\rmn{T}}\hphantom{0} $}& \multicolumn{1}{c}{$\eta$}\\
 &\multicolumn{1}{c}{$\rmn{kpc}$}	&	&\multicolumn{1}{c}{$\rmn{keV}$}	&\multicolumn{1}{c}{$\rmn{keV}$} 	& \multicolumn{1}{c}{$\rmn{kpc}$\hphantom{0}}\\\hline
Centaurus	&	63	&	0.2\hphantom{0}	&	0.8	&	5.1	&	21\hphantom{$^{\rmn{(3)}}$}	&	1.0	\\
Hydra A	&	297		&	0.5\hphantom{0}	&	2.5	&	5.9	&	300$^{\rmn{(3)}}$	&	0.5	\\
Virgo	&	54		&	0.2\hphantom{0}	&	1.9	&	3.1	&	28\hphantom{$^{\rmn{(3)}}$}	&	1.4	\\
A 85	&	248		&	0.3\hphantom{0}	&	3.0	&	8.7	&	92\hphantom{$^{\rmn{(3)}}$}	&	1.2	\\
A 133	&	136		&	0.2\hphantom{0}	&	2.3	&	4.8	&	51\hphantom{$^{\rmn{(3)}}$}	&	2.6	\\
A 262	&	81		&	0.2\hphantom{0}	&	1.5	&	2.5	&	9\hphantom{$^{\rmn{(3)}}$}	&	2.5	\\
A 383	&	289		&	0.2\hphantom{0}	&	3.0	&	5.9	&	57\hphantom{$^{\rmn{(3)}}$}	&	3.0	\\
A 496	&	79		&	0.1\hphantom{0}	&	1.9	&	24.4&	390\hphantom{$^{\rmn{(3)}}$}	&	1.0	\\
A 539	&	146		&	0.2\hphantom{0}	&	3.0	&	3.2	&	21\hphantom{$^{\rmn{(3)}}$}	&	10.0	\\
A 907	&	635		&	0.2\hphantom{0}	&	3.5	&	6.5	&	47\hphantom{$^{\rmn{(3)}}$}	&	1.7	\\
A 1644	&	292		&	0.2\hphantom{0}	&	2.0	&	5.5	&	49\hphantom{$^{\rmn{(3)}}$}	&	1.5	\\
A 1795	&	377		&	0.2\hphantom{0}	&	3.3	&	7.6	&	75\hphantom{$^{\rmn{(3)}}$}	&	1.5	\\
A 1991	&	197		&	0.2\hphantom{0}	&	0.9	&	3.0	&	19\hphantom{$^{\rmn{(3)}}$}	&	1.5	\\
A 2052	&	122		&	0.2\hphantom{0}	&	1.5	&	3.5	&	23\hphantom{$^{\rmn{(3)}}$}	&	2.8	\\
A 2199	&	104		&	0.1\hphantom{0}	&	2.7	&	4.9	&	24\hphantom{$^{\rmn{(3)}}$}	&	2.1	\\
A 2597	&	87		&	0.1\hphantom{0}	&	2.3	&	5.0	&	41\hphantom{$^{\rmn{(3)}}$}	&	1.6	\\
A 3112	&	245		&	0.2\hphantom{0}	&	2.7	&	5.7	&	35\hphantom{$^{\rmn{(3)}}$}	&	1.6	\\
A 3581	&	107		&	0.2\hphantom{0}	&	1.4	&	4.2	&	115\hphantom{$^{\rmn{(3)}}$}	&	1.4	\\
A 4059	&	221		&	0.2\hphantom{0}	&	2.1	&	5.0	&	30\hphantom{$^{\rmn{(3)}}$}	&	2.0	\\
AWM 7	&	78		&	0.2\hphantom{0}	&	2.6	&	3.8	&	14\hphantom{$^{\rmn{(3)}}$}	&	2.5	\\
MKW 3S&	229		&	0.1\hphantom{0}	&	3.1	&	3.7	&	26\hphantom{$^{\rmn{(3)}}$}	&	3.3	\\
MKW 4	&	48		&	0.2\hphantom{0}	&	1.5	&	2.2	&	10\hphantom{$^{\rmn{(3)}}$}	&	3.4	\\
PKS 0745&	416		&	0.5\hphantom{0}	&	3.2	&	20.0	&	300$^{\rmn{(3)}}$	&	1.0	\\
ZwCl 1742&	343		&	0.2\hphantom{0}	&	3.0	&	4.3	&	51\hphantom{$^{\rmn{(3)}}$}	&	6.1	\\
Ophiuchus	&	257	&	0.2\hphantom{0}	&	0.8	&	9.3	&	14\hphantom{$^{\rmn{(3)}}$}	&	1.1	\\
Perseus	&	115		&	0.2\hphantom{0}	&	3.2	&	8.7	&	94\hphantom{$^{\rmn{(3)}}$}	&	2.3	\\
2A 0335	&	148	&	0.2\hphantom{0}	&	1.6	&	4.9	&	46\hphantom{$^{\rmn{(3)}}$}	&	1.8	\\
A 478	&	444		&	0.5\hphantom{0}	&	3.0	&	6.8	&	26\hphantom{$^{\rmn{(3)}}$}	&	1.9	\\
A 1835	&	590		&	0.4\hphantom{0}	&	2.6	&	17.	&	166\hphantom{$^{\rmn{(3)}}$}	&	0.7\\
A 2029	&	497		&	0.2\hphantom{0}	&	1.7	&	14.7&	121\hphantom{$^{\rmn{(3)}}$}	&	0.4	\\
A 2204	&	1040	&	0.2\hphantom{0}	&	3.3	&	10.2	&	40\hphantom{$^{\rmn{(3)}}$}	&	3.0	\\
A 2390	&	549		&	0.3\hphantom{0}	&	4.0	&	12.7	&	52\hphantom{$^{\rmn{(3)}}$}	&	1.6	\\
MS 1455	&	486	&	0.2\hphantom{0}	&	1.5	&	7.7	&	50$^{\rmn{(3)}}$	&	0.4	\\
RBS 797	&	315	&	0.5\hphantom{0}	&	4.2	&	15.2&	217\hphantom{$^{\rmn{(3)}}$}	&	1.2	\\
RX J1347	&	501	&	0.08	&	6.6	&	23.8	&	82\hphantom{$^{\rmn{(3)}}$}	&	1.5	\\
RX J1504	&	587	&	0.5\hphantom{0}	&	4.4	&	10.9	&	105\hphantom{$^{\rmn{(3)}}$}	&	1.4	\\
RX J1532	&	477	&	0.4\hphantom{0}	&	4.1	&	7.6	&	108	\hphantom{$^{\rmn{(3)}}$}&	2.0	\\
RX J1720	&	367	&	0.4\hphantom{0}	&	4.4	&	7.5	&	86\hphantom{$^{\rmn{(3)}}$}	&	2.7	\\
ZwCl 3146	&	382	&	0.3\hphantom{0}	&	3.7	&	8.7	&	63\hphantom{$^{\rmn{(3)}}$}	&	2.3	\\
\hline
\end{tabular}
\end{threeparttable}
\begin{tablenotes}
\item (1) Maximal radius that we include in fit.
\item (2) Parameter fixed in fit.
\item (3) Fixed value for $r_\rmn{T}$.
\end{tablenotes}
\end{table}

\bsp	
\label{lastpage}
\end{document}